\documentclass[12pt]{article}
\usepackage{amsmath}
\usepackage{graphicx,psfrag,epsf}
\usepackage{enumerate}
\usepackage{natbib}

\newcommand{\blind}{0}

\usepackage{subcaption} 
\usepackage{arydshln}
\usepackage{multirow}
\usepackage{amsmath}
\usepackage{amssymb}
\usepackage{arydshln}
\usepackage{amsfonts}
\usepackage{algorithm}
\usepackage{algorithmic}
\usepackage{caption}

\usepackage{subfloat}
\usepackage{ntheorem}
\usepackage{enumerate}
\usepackage{verbatim}
\usepackage{graphicx}
\newtheorem{theorem}{Theorem}

\newtheorem{proposition}{Proposition}

\newtheorem{assumption}{Assumption}

\usepackage[colorlinks=true, allcolors=blue]{hyperref}
\usepackage{color}
\usepackage{multirow}
\definecolor{grey}{RGB}{80,96,150}

\newcounter{xxx}
\newcommand{\xbold}{\boldsymbol{x}}
\newcommand{\ybold}{\boldsymbol{y}}

\newcommand{\thetabold}{{\boldsymbol{\theta}}}
\newcommand{\eg}{\emph{e.g.}}

\usepackage{lscape}

\begin{document}

\def\spacingset#1{\renewcommand{\baselinestretch}%
{#1}\small\normalsize} \spacingset{1}


\if0\blind
{
  \title{\bf A multifidelity approximate Bayesian computation with pre-filtering}
   \author{Xuefei Cao$^1$, 
    Shijia Wang$^{2*}$ 
    and
    Yongdao Zhou$^1$\thanks{
    Address correspondence to: Dr.\ Shijia Wang ({\tt wangshj1@shanghaitech.edu.cn}) and Dr.\ Yongdao Zhou ({\tt ydzhou@nankai.edu.cn}).}\\
    $^1$School of Statistics and Data Science, Nankai University, China\\
    $^2$Institute of Mathematical Sciences, ShanghaiTech University, China\\
    }
  \maketitle

} \fi

\spacingset{1.5}
\begin{abstract}
    
Approximate Bayesian Computation (ABC) methods often require extensive simulations, resulting in high computational costs. This paper focuses on multifidelity simulation models and proposes a pre-filtering hierarchical importance sampling algorithm. 
Under mild assumptions, we theoretically prove that the proposed algorithm satisfies posterior concentration properties, characterize the error upper bound and the relationship between algorithmic efficiency and pre-filtering criteria. Additionally, we provide a practical strategy to assess the suitability of multifidelity models for the proposed method. 
Finally, we develop a multifidelity ABC sequential Monte Carlo with adaptive pre-filtering strategy. 
Numerical experiments are used to demonstrate the effectiveness of the proposed approach. We develop an R package that is available at \url{https://github.com/caofff/MAPS}.

\end{abstract}
\section{Introduction}
Bayesian inference provides a principled framework for parameter estimation and uncertainty quantification by combining prior knowledge with observed data through the likelihood function. However, for many complex models encountered in fields such as systems biology, epidemiology, and cosmology, direct evaluation of the likelihood can be computationally prohibitive or infeasible. This challenge has led to the development of likelihood-free inference techniques \citep{papamakarios2016fast, papamakarios2019sequential, hermans2020likelihood, boelts2024sbi}, among which Approximate Bayesian Computation (ABC) \citep{Pritchard1999Population, Beaumont2002ABC, sisson2018handbook} is a prominent method. ABC avoids explicit likelihood calculations by simulating data from the model and comparing these simulated datasets to observed data, thereby providing an effective approach for parameter inference in complex settings.

The accuracy of ABC posterior estimates depends heavily on extensive sampling and simulation throughout the parameter space. To overcome this inefficiency, various algorithms have been developed that explore the parameter space more effectively by designing tailored proposal distributions, thus reducing simulations in regions with low posterior probability. For example, ABC Markov Chain Monte Carlo (ABC-MCMC) \citep{Marjoram2003MCMC, Wegmann2009MCMC} generates samples via a Markov chain whose stationary distribution is the target posterior, improving efficiency by proposing new parameters near the current particle. Importance sampling methods correct the mismatch between the proposal and  target distributions, and the algorithmic performance relies on the closeness of two distributions. ABC sequential Monte Carlo (ABC-SMC) \citep{Sisson2007SMC, del2012adaptive} is an extension of importance sampling, which propagates samples through a sequence of intermediate distributions defined by a decreasing series of tolerance thresholds, utilizing importance weighting and resampling to approximate the posterior.

To address the challenge of high cost of expensive simulations in ABC, one research line involves the usage of multi-fidelity algorithms, which attempt to accept or reject parameter proposals early by leveraging cheaper, lower-fidelity simulations before performing expensive high-fidelity (HF) runs. 
For instance, \cite{Picchini2014SDE} proposed an early-rejection ABC-MCMC algorithm specifically for parameter estimation in stochastic differential equations, decomposing the Metropolis-Hastings acceptance ratio into two stages to save computational resources.  \cite{Everitt2021DASMC} integrated delayed acceptance (DA) MCMC \citep{christen2005markov} within the ABC framework to first screen samples using inexpensive approximations. Furthermore, \cite{cao2025using} reduced ABC-MCMC simulation costs by developing a Gaussian process model to estimate simulation discrepancy and reject unlikely parameters prior to full simulations.

In many real applications, simulation models often have multiple fidelity levels due to complexity and computational costs. Incorporating multifidelity models into approximate Bayesian inference is essential to balance computing speed and accuracy. Low-fidelity (LF) models enable fast but approximate evaluations, while HF models provide precise but costly results. Ignoring this hierarchy can lead to excessive computation or poor inference. Hence, multifidelity methods is crucial for scalable and reliable Bayesian inference of complex applications. The multi-fidelity approximate Bayesian computation (MF-ABC) framework proposed by \cite{prescott2020multifidelity} enables both early acceptance and early rejection by estimating the likelihood through a combination of LF and HF model simulations. Building on this framework, \cite{prescott2021multifidelity} integrated MF-ABC with sequential Monte Carlo (SMC) techniques. Moreover, \cite{krouglova2025multifidelity} proposed a multifidelity transfer-learning framework that pre-trains neural posterior estimators using computationally efficient LF simulations and subsequently fine-tunes them with limited HF data. 

Other approaches employ variance reduction techniques, using LF estimates to decrease the variance of the estimator. \citet{warne2018multilevel} accelerate ABC inference by leveraging multifidelity simulations and directly applying multilevel Monte Carlo variance reduction techniques to rejection sampling. Multilevel SMC \citep{jasra2019multilevel} bypasses the need for explicit level couplings by utilizing an SMC-based importance sampling scheme, reusing particles across tolerance levels and exploiting the multilevel identity to achieve accurate ABC inference with reduced computational cost. 

In this paper, we consider simulation models with outputs of varying fidelity levels, where HF simulations provide greater accuracy at a higher computational cost, while LF simulations are less accurate but computationally cheaper. For example, in stochastic differential equations, a smaller time step yields more precise results but incurs higher costs, whereas a larger time step reduces accuracy but accelerates simulation and lowers computational expenses. We propose a pre-filtering stratified importance sampling method that employs LF simulations to filter out parameters unlikely to belong to the posterior before conducting HF simulations, thereby reducing the need to generate costly HF data. We theoretically establish that, under mild assumptions, the posterior distribution obtained via the pre-filtering stratified importance sampling satisfies posterior concentration properties. Furthermore, we derive explicit relationships between the algorithm’s accuracy, efficiency, and the pre-filtering criteria, revealing a trade-off between filtering conditions and estimation precision. We also provide practical strategies to assess the suitability of multifidelity models for the proposed algorithm. Building upon the adaptive sequential Monte Carlo (ASMC) framework introduced by \cite{del2012adaptive}, we further develop an adaptive pre-filtering SMC algorithm that dynamically selects filtering thresholds throughout the iterative process, balancing accuracy and computational efficiency. Numerical experiments demonstrate that our method reduces the number of HF simulations by approximately 40\% compared to the standard ASMC method.

This paper is organized as follows: Section \ref{sec:method} reviews the fundamentals of ABC and SMC algorithms. Section \ref{sec:mfabc} introduces the proposed multifidelity ABC model, along with its properties, and presents a strategy to assess the suitability of multifidelity models. Section \ref{sec:maps} describes the adaptive pre-screening sequential Monte Carlo algorithm. Section \ref{sec:numerical} presents numerical experiments, and Section \ref{sec: con} concludes the paper.
We provide a summary of notations used in this paper in Appendix S.4.

\section{Preliminaries}
\label{sec:method}
In this section, we provide a review of ABC and the adaptive sequential Monte Carlo (ABC-SMC) algorithm, which form the foundation for the pre-filtering importance sampling framework applied to multifidelity simulators.
\subsection{Approximate Bayesian computation}
We consider a simulation model $ P_{\boldsymbol{\theta}} $ with input parameter $\boldsymbol{\theta}\in\Theta$ and output random variable $\boldsymbol{x}\in \mathcal{X}$.  The likelihood function $ p(\boldsymbol{x} \mid \boldsymbol{\theta}) $ is intractable, but we could generate simulated data from the model as $\boldsymbol{x} \sim p(\cdot \mid \boldsymbol{\theta}) $ or $ \boldsymbol{x} \sim P_{\boldsymbol{\theta}}$.

Given observed data $\boldsymbol{y}$ and prior distribution $\pi(\boldsymbol{\theta})$, the ABC posterior distribution is defined as
\begin{equation}
\label{eq:abc}
    \pi_\varepsilon \equiv \pi_{\varepsilon}(\boldsymbol{\theta} \mid \boldsymbol{y}) \propto \pi(\boldsymbol{\theta}) p_\varepsilon(\thetabold\mid\ybold),
\end{equation}
with 
\begin{equation}
\label{eq:HF-likelihood}
p_\varepsilon(\thetabold\mid\ybold)=\int \mathbb{I}\big(\Delta(\boldsymbol{x}, \boldsymbol{y}) \leq \varepsilon\big) \, p(\boldsymbol{x} \mid \boldsymbol{\theta}) \, d\boldsymbol{x},\end{equation} 
where $\varepsilon > 0$ is a tolerance threshold, 
$\mathbb{I}(\cdot)$ is an indicator function that equals 1 if the condition holds and 0 otherwise, and  $\Delta(\xbold,\ybold)$ denotes the discrepancy between $\xbold$ and $\ybold$. For simplicity, and without causing confusion, we denote \(p_\varepsilon(\boldsymbol{\theta} \mid \boldsymbol{y})\) simply as \(p_\varepsilon(\boldsymbol{\theta})\) in the sequel. This means that only parameter values generating simulated data sufficiently close to the observed data (within tolerance $\varepsilon$) are accepted, providing an approximation to the true posterior. As $\varepsilon \to 0$, the ABC posterior converges to the exact Bayesian posterior. The integral in the Eq.\eqref{eq:HF-likelihood} can be interpreted as an approximation of the likelihood $ p(\boldsymbol{y}_{obs} \mid \boldsymbol{\theta}) $. 

In practice, $\Delta(\cdot,\cdot)=d(S(\cdot),S(\cdot))$, where $d(\cdot,\cdot)$ is a distance function. Summary statistics $S(\cdot)$ are commonly employed when the observed data $\boldsymbol{y}$ and simulated data $\boldsymbol{x}$ are high-dimensional or structurally complex. The use of summary statistics serves to reduce data dimensionality, thereby facilitating more efficient and tractable comparisons between simulated and observed data within the ABC framework. Ideally, these summary statistics should retain as much information about the model parameters as possible. The selection and construction of informative summary statistics remains an active area of research \citep{sisson2011likelihood, marin2012approximate, blum2013comparative}. In this paper, we do not address the construction of summary statistics and assume they are given. The integral of Eq.\eqref{eq:abc} is approximated using Monte Carlo (MC) sampling, that is,
\begin{equation}
\label{eq:likelihood}
\hat{p}_\varepsilon( \boldsymbol{\theta}) = \frac{1}{n} \sum_{i=1}^n \mathbb{I}\big( \Delta(\boldsymbol{x}_i, \boldsymbol{y}_{obs}) \leq \varepsilon \big),
\end{equation} where $\boldsymbol{x}_i \sim p(\cdot \mid \boldsymbol{\theta})$ for $i=1,\ldots,n$. A larger number of simulated data $n$ provides more accurate estimate of the likelihood function $p(\ybold\mid\thetabold)$.

\subsection{An adaptive ABC sequence Monte Carlo algorithm}

The simplest ABC algorithm employs basic rejection sampling. However, when there is a significant discrepancy between the prior and posterior distributions, the efficiency of this method becomes extremely low. Although more effective importance sampling schemes exist, it is often challenging to design high-quality proposal distributions. Commonly used approaches, such as ABC-MCMC, tend to get trapped in local regions and exhibit slow convergence rates, particularly when the tolerance 
$\varepsilon$ is set to a small value. The ABC Sequential Monte Carlo (ABC-SMC) \citep{Sisson2007SMC} algorithm mitigates these issues by introducing a sequence of intermediate target distributions with decreasing thresholds. By combining importance sampling and ABC-MCMC, ABC-SMC alleviates the difficulties of constructing suitable proposal distributions and the slow convergence associated with small threshold values in ABC-MCMC.  Furthermore, the adaptive sequential Monte Carlo algorithm (ABC-ASMC) \citep{del2012adaptive} proposes an adaptive strategy for determining the decreasing threshold sequence. In the following, we provide a detailed description of the ABC-ASMC algorithm.

The ABC-SMC method employs a sequence of decreasing tolerance thresholds $\{\varepsilon_0, \ldots, \varepsilon_T\}$,  with $\varepsilon_0 = \infty$ and $\varepsilon_T = \varepsilon$. The  corresponding intermediate distributions are denoted by $\pi_{\varepsilon_t}(\boldsymbol{\theta} \mid \boldsymbol{y})$. At each iteration $t$, the importance proposal is constructed using the target distribution from the previous iteration, $\pi_{\varepsilon_{t-1}}(\boldsymbol{\theta} \mid \boldsymbol{y})$, and new samples are drawn accordingly. This iterative process ensures that the weighted samples progressively approximate the target posterior distribution, reducing the discrepancy between the proposal and target distributions, and thereby improving sampling efficiency.

Let $\{(\boldsymbol{\theta}_{t-1}^{(i)}, W_{t-1}^{(i)})\}_{i=1}^N$ denote the weighted particle set at iteration $t-1$. In the adaptive SMC framework, the tolerance $\varepsilon_t$ is chosen such that a fixed proportion $\alpha$ of “active” particles from the previous iteration is preserved. Specifically, $\varepsilon_t$ is set to satisfy
$$
PA(\{W_{t}^{(i)}\}; \varepsilon_t) = \alpha \, PA(\{W_{t-1}^{(i)}\}),
$$
where $PA(\{W_{t}^{(i)}\}; \varepsilon_t) = \frac{1}{N} \sum_{i=1}^N \mathbb{I}\{ W_{t}^{(i)} > 0 \}$ represents the proportion of particles with positive weights at tolerance $\varepsilon_t$, and $PA(\{W_{t-1}^{(i)}\})= \frac{1}{N} \sum_{i=1}^N \mathbb{I}\{ W_{t-1}^{(i)} > 0 \}$ where $\left\{W_{t-1}^{(i)}\right\}_{i=1:N}$ is given. Here, $\mathbb{I}\{\cdot\}$ denotes the indicator function. This adaptive selection of $\varepsilon_t$ maintains particle diversity while gradually tightening the posterior approximation. The particle weights are then updated as follows:
\begin{equation}
\label{eq:weight update}
W_t^{(i)} \propto W_{t-1}^{(i)} \frac{\sum_k \mathbb{I}\big(\Delta(\xbold_{t-1,k}^{(i)}, \ybold) < \varepsilon_t\big)}{\sum_k \mathbb{I}\big(\Delta(\xbold_{t-1,k}^{(i)}, \ybold) < \varepsilon_{t-1}\big)}.
\end{equation}

Then we conduct a resampling step to stabilize the algorithm and mitigate particle degeneracy. All particles are equally weighted after resampling.

To improve exploration of the parameter space, a Markov kernel targeting $\pi_{\varepsilon_t}$ is applied, typically implemented via a Metropolis-Hastings (MH) step. Given the current state $(\thetabold, \xbold_{1:n})$, a candidate $(\thetabold^*, \xbold_{1:n}^*)$ is proposed from $q_t(\thetabold^*\mid\thetabold) \prod_{k=1}^n p(\xbold_k^* \mid \thetabold^*)$. The proposal is accepted with probability
\begin{equation}
\label{eq:MH acceptance ratio}
\alpha( (\thetabold,\xbold_{1:n}),(\thetabold^*,\xbold^*_{1:n})) = \min\left\{1, \frac{\pi(\thetabold^*) \sum_{k=1}^n \mathbb{I}(\Delta(\xbold^*_k,\ybold)<\varepsilon_t)}{\pi(\thetabold) \sum_{k=1}^n \mathbb{I}(\Delta(\xbold_k,\ybold)<\varepsilon_t)}\frac{q_t(\thetabold\mid\thetabold^*)}{q_t(\thetabold^*\mid \thetabold)}\right\}.
\end{equation}

The details of ABC-ASMC are shown in Supplement Algorithm S.1.

\section{Pre-filtering  hierarchical importance sampling for multi-fidelity simulation model}
\label{sec:mfabc}

 In this section, we propose a pre-filtering algorithm tailored for multi-fidelity simulation models. The section begins by introducing the multi-fidelity simulation framework and outlining its key principles. We then present a pre-filtering hierarchical importance sampling method, which leverages hierarchical fidelity levels to improve sampling efficiency. Subsequently, an adaptive pre-filtering SMC algorithm is introduced, which adaptively utilizes low-fidelity (LF) simulations for pre-filtering in order to reduce the number of high-fidelity (HF) simulations required. Finally, the theoretical properties of the proposed algorithm are discussed. 
\subsection{A multi-fidelity simulator}
In multi-fidelity frameworks, simulation models of varying fidelity levels are employed. The HF model generates simulation data with high accuracy but at a significant computational cost. In contrast, the LF model produces less accurate data but requires substantially less computational effort. Such LF models may arise naturally, for instance, through coarser time discretizations in differential equation solvers, or may be constructed as surrogate models. The development of these LF models is outside the scope of this paper. The conditions that requires a LF simulations to satisfy will be presented in the following sections. 

Within the multi-fidelity ABC framework, the primary model $ P_{\boldsymbol{\theta}} $ is characterized by the likelihood function $ p(\cdot \mid \boldsymbol{\theta}) $, corresponding to the HF simulator. The LF simulator is denoted by $ \tilde{P}_{\boldsymbol{\theta}} $, associated with the likelihood $ \tilde{p}(\cdot \mid \boldsymbol{\theta}) $. For any parameter $\boldsymbol{\theta} \in \Theta$, simulation outputs $\boldsymbol{x} \sim p(\cdot \mid \boldsymbol{\theta})$ and $\tilde{\xbold} \sim \tilde{p}(\cdot \mid \boldsymbol{\theta})$ can be generated independently. The LF ABC posterior distribution is
\begin{equation}
    \tilde{\pi}_{\tilde{\varepsilon}}(\boldsymbol{\theta} \mid \boldsymbol{y}) \propto \pi(\boldsymbol{\theta}) \tilde{p}_{\tilde{\varepsilon}}(\boldsymbol{\theta} \mid \boldsymbol{y}),
    \label{eq:ABC-low}
\end{equation}
where the approximated LF likelihood is defined as
\[
\tilde{p}_{\tilde{\varepsilon}}(\boldsymbol{\theta} \mid \boldsymbol{y}) = \int \mathbb{I}(\Delta(\tilde{\boldsymbol{x}}, \boldsymbol{y}) < \tilde{\varepsilon}) \tilde{p}(\tilde{\boldsymbol{x}} \mid \boldsymbol{\theta}) d\tilde{\boldsymbol{x}}.
\]
For simplicity,  we denote \(\tilde{p}_{\tilde{\varepsilon}}(\boldsymbol{\theta} \mid \boldsymbol{y})\) simply as \(\tilde{p}_{\tilde{\varepsilon}}(\boldsymbol{\theta})\) in the sequel.

We assume that the observed data $\boldsymbol{y}$ can be mapped into the LF data space via a mapping $ M(\cdot) $, such that $\tilde{\ybold} = M(\boldsymbol{y})$. This mapping allows for the evaluation of discrepancies between LF simulations and observations through the discrepancy function $\Delta(\tilde{\xbold}, \tilde{\ybold})$. For example, if $\boldsymbol{y}$ represents a time series, $\tilde{\ybold}$ may correspond to a sparser version of the time series, and $M$ represents the downsampling operation. In this study, we further simplify by assuming that the data spaces coincide, i.e., $\boldsymbol{y} = \tilde{\ybold}$, and that the discrepancy functions satisfy
$
\Delta(\tilde{\xbold}, \boldsymbol{y}) = \Delta(\tilde{\xbold}, \tilde{\ybold}).
$

 \subsection{A pre-filtering hierarchical importance sampling}
In ABC importance sampling algorithms, using the prior distribution as the importance proposal is straightforward and intuitive. However, when there is a significant discrepancy between the prior and the posterior distributions, this approach tends to be inefficient, often resulting in a low acceptance rate and many wasted simulations. To improve efficiency, it is desirable to design an importance proposal distribution that closely approximates the ABC posterior. A well-chosen proposal can generate a higher proportion of effective samples with a fixed computational budget, which is especially beneficial when dealing with computationally expensive simulators. Nonetheless, constructing such an importance proposal distribution is a challenging task, with complexity comparable to directly performing inference on the ABC posterior. 

In this paper, we exploit the consistency between LF and HF simulations by leveraging the computationally inexpensive LF ABC posterior to improve the importance proposal distribution. The detailed algorithmic procedure is summarized in Algorithm \ref{alg:MFABC-IS}. Initially, parameter samples \(\boldsymbol{\theta}_{1:N}\) are drawn from the importance proposal distribution \(q(\cdot)\), LF simulations are conducted, and the corresponding LF weights are assigned:
\begin{equation}
    \label{eq:ISIRweight_LF}
    W^{(i)} = \frac{\pi(\boldsymbol{\theta}^{(i)})}{q(\boldsymbol{\theta}^{(i)})} \sum_{k=1}^{n_L} \mathbb{I}\left(\Delta(\tilde{\boldsymbol{x}}^{(i)}_k, \boldsymbol{y}) < \tilde{\varepsilon}\right),
\end{equation}
where \(\tilde{\boldsymbol{x}}^{(i)}_k\) denotes the \(k\)th LF simulation output for the \(i\)th parameter sample, and \(\tilde{\varepsilon}\) is the corresponding LF tolerance threshold. For samples with \(W^{(i)} > 0\), HF simulations are further performed to update the weights:
\begin{equation}
    \label{eq:ISIRweight_HF}
    W^{(i)} = W^{(i)} \cdot \frac{\sum_{k=1}^{n_H} \mathbb{I}\left(\Delta(\boldsymbol{x}^{(i)}_k, \boldsymbol{y}) < \varepsilon\right)}{\sum_{k=1}^{n_L} \mathbb{I}\left(\Delta(\tilde{\boldsymbol{x}}^{(i)}_k, \boldsymbol{y}) < \tilde{\varepsilon}\right)} = \frac{\pi(\boldsymbol{\theta}^{(i)})}{q(\boldsymbol{\theta}^{(i)})}\sum_{k=1}^{n_H} \mathbb{I}\left(\Delta(\boldsymbol{x}^{(i)}_k, \boldsymbol{y}) < \varepsilon\right),
\end{equation}
where \(\boldsymbol{x}^{(i)}_k\) represents the \(k\)th HF simulation output for the \(i\)th parameter, and \(\varepsilon\) is the HF tolerance threshold.
\begin{algorithm}[H]
\caption{A pre-filtering hierarchical importance sampling}
\label{alg:MFABC-IS}

\textbf{Input:} Prior \(\pi(\cdot)\), importance proposal distribution \(q(\cdot)\), LF simulator \(\tilde{p}(\cdot \mid\boldsymbol{\theta})\), HF simulator \(p(\cdot \mid\boldsymbol{\theta})\), observed data \(\boldsymbol{y}\), tolerance thresholds \(\varepsilon, \tilde{\varepsilon}\), number of samples \(N\), number of LF simulations \(n_L\), number of HF simulations \(n_H\).

\textbf{Output:} Weighted samples \(\{(W^{(i)}, \boldsymbol{\theta}^{(i)})\}_{i=1}^N\).

\begin{algorithmic}[1]
\FOR{$i=1,\ldots,N$}
\STATE Draw \(\boldsymbol{\theta}^{(i)} \sim q(\cdot)\).
\STATE Generate \(n_L\) LF simulations \(\tilde{\boldsymbol{x}}^{(i)}_{1:n_L} \sim \tilde{p}(\cdot \mid \boldsymbol{\theta}^{(i)})\), and compute \(W^{(i)}\) using Eq. \eqref{eq:ISIRweight_LF}.
\IF{$W^{(i)} > 0$}
\STATE Generate \(n_H\) HF simulations \(\boldsymbol{x}^{(i)}_{1:n_H} \sim p(\cdot \mid \boldsymbol{\theta}^{(i)})\), and update \(W^{(i)}\) according to Eq. \eqref{eq:ISIRweight_HF}.
\ENDIF
\ENDFOR
\STATE Normalize the weights \(\{W^{(i)}\}_{i=1}^N\).
\end{algorithmic}
\end{algorithm}

The pre-filtering strategy in Algorithm \ref{alg:MFABC-IS} relies solely on LF simulations to screen out samples unlikely to contribute significantly before conducting computationally expensive HF simulations. Importantly, this approach does not employ the LF ABC posterior distribution \(\tilde{\pi}_{\tilde{\varepsilon}}\) to construct a full importance proposal distribution. The reason is that \(\tilde{\pi}_{\tilde{\varepsilon}}\) is typically estimated via Monte Carlo methods, which leads to high variance in the importance weights \(\pi_{\varepsilon} / \tilde{\pi}_{\tilde{\varepsilon}}\), as \(\tilde{\pi}_{\tilde{\varepsilon}}\) appears in the denominator. Such variance adversely affects estimation accuracy and overall algorithmic performance. Algorithm \ref{alg:MFABC-IS} effectively harnesses LF simulations to enhance the proposal while avoiding this pitfall.

In Algorithm \ref{alg:MFABC-IS}, the weights of parameter samples \(\{\boldsymbol{\theta}_i\}_{i=1}^N\) for the pre-filtering hierarchical importance sampling are
\[
W^{(i)} \propto \frac{\pi(\boldsymbol{\theta}^{(i)})}{q(\boldsymbol{\theta}^{(i)})}\mathbb{I}\left(\min_k \Delta(\tilde{\boldsymbol{x}}^{(i)}_k, \boldsymbol{y}) < \tilde{\varepsilon}\right) \cdot \sum_{k=1}^{n_H} \mathbb{I}\left(\Delta(\boldsymbol{x}^{(i)}_k, \boldsymbol{y}) < \varepsilon\right).
\] This formulation implies that introducing the LF pre-filtering step induces an estimation bias. The following subsection presents a theoretical justification for this point.
\subsection{Theoretical  property}
In this section, we show some properties of the proposed multifidelity simulator. 
Proposition \ref{prop:ISIR-posterior} presents the approximate posterior distribution produced by Algorithm \ref{alg:MFABC-IS}, and Proposition \ref{prop:ISR-ACC} analyzes the impact of LF pre-filtering on both the accuracy of the estimates and the computational efficiency.  Theorem \ref{th:concentration} demonstrates the posterior concentration of the proposed algorithm.

\begin{proposition}
    The weighted samples obtained by running Algorithm \ref{alg:MFABC-IS} approximate the multi-fidelity ABC posterior, which is defined as
\begin{equation}
    \label{eq: MAPS-posterior}\pi_{\varepsilon,\tilde{\varepsilon}}\equiv
\pi_{\varepsilon,\tilde{\varepsilon}}(\boldsymbol{\theta} \mid \boldsymbol{y}) \propto \pi(\boldsymbol{\theta}) \, p_{\varepsilon}( \boldsymbol{\theta}) \, {Q}_{\tilde{\varepsilon}}(\boldsymbol{\theta}),
\end{equation}
where
\[
{Q}_{\tilde{\varepsilon}}(\boldsymbol{\theta}) = 1 - \left(1 - \tilde{p}_{\tilde{\varepsilon}}(\boldsymbol{\theta})\right)^{n_L},
\]
and the parameter \(n_L\) represents the number of LF simulations. 
\label{prop:ISIR-posterior}
\end{proposition}

\begin{assumption}
\label{Ass:1}
For any \(\varepsilon > 0\), there exists a \(\tilde{\varepsilon}>0\) and constant $a_0\in(0,1)$ such that  
$a_L < 1$,
where  
\begin{equation}
\label{eq:a_L}
a_L = \int \pi_\varepsilon(\boldsymbol{\theta}) \left(1 - \tilde{p}_{\tilde{\varepsilon}}(\boldsymbol{\theta})\right)^{n_L} d\boldsymbol{\theta}.
\end{equation}
\end{assumption}
Assumption \ref{Ass:1} regulates the fraction of posterior samples excluded through the pre-filtering procedure, thereby providing an essential guarantee that the output set of the algorithm is non-empty.

\begin{theorem}
\label{th:concentration}
     If  Assumptions 1-3 of \cite{frazier2018asymptotic} and Assumption \ref{Ass:1} and S.4 stated in this paper are satisfied, the posterior concentration of the pre-filtering hierarchical importance sampling targeting  \eqref{eq: MAPS-posterior}  still holds.  
\end{theorem}

Theorem \ref{th:concentration} demonstrates that, under reasonable assumptions on the LF simulator, the proposed algorithm retains posterior concentration. This property is critical since, for any subset \( A \subset \Theta \), the ABC posterior probability \(\Pi_{\varepsilon}\{A \mid \ybold\}\) generally deviates from the exact posterior, where \(\Pi_\varepsilon(\cdot\mid\ybold)\) denotes the ABC posterior measure with density \(\pi_\varepsilon(\cdot\mid\ybold)\). In the absence of exact posterior guarantees, the concentration of \(\Pi_{\varepsilon}\{\cdot \mid \ybold\}\) around the true parameter \(\boldsymbol{\theta}_0\) provides a robust basis for quantifying uncertainty in \(\boldsymbol{\theta}\). Detailed technical discussions and proofs are provided in the Appendix S.2.

\begin{proposition}
\begin{enumerate}[(i)]
    \item For any \(\varepsilon\) and \(\tilde{\varepsilon}\) such that \(a_L < 1\) holds, the approximation error introduced by incorporating a LF filtering step within the pre-filtering hierarchical importance sampling framework is bounded by
    \begin{equation}
        \left\| \pi_{\varepsilon,\tilde{\varepsilon}} - \pi_\varepsilon \right\|_{1} < \frac{1}{1-a_L} - (1-a_L).
    \end{equation}

\label{prop:ISR-ACC-acc}
\item The expected pre-filtering rate of the hierarchical importance sampling algorithm is given by \label{prop:ISR-ACC-eff}
\begin{equation}
    1-\alpha_L=\int q(\thetabold)(1-Q_\varepsilon(\thetabold))d\thetabold.
    \label{eq:alpha_L},
\end{equation}
where $q(\thetabold)$ is the important proposal distribution, such that $q(\thetabold)=\pi(\thetabold)$.
\end{enumerate}
\label{prop:ISR-ACC}
\end{proposition}

Proposition \ref{prop:ISR-ACC} \eqref{prop:ISR-ACC-acc} provides an explicit upper bound on the \(L_1\) distance between the joint posterior distribution \(\pi_{\varepsilon,\tilde{\varepsilon}}\) obtained from the two-stage LF and HF ABC approach and the original ABC posterior \(\pi_\varepsilon\). Proposition \ref{prop:ISR-ACC}  \eqref{prop:ISR-ACC-eff}  gives the pre-filtering rate at the LF level, which reflects the proportion of parameter samples that can be discarded early, thereby avoiding the computational cost of expensive HF simulations.

This result offers a theoretical guarantee for using LF models as a computationally inexpensive screening step: although the LF model introduces certain bias by altering the sample distribution through an imperfect filtering mechanism, the resulting error is controllable and bounded. Such error bounds ensure that multifidelity ABC methods reduce computational cost while maintaining reasonable accuracy. 

The false rejection rate \(a_L\) is supposed to be as small as possible, and the pre-filtering rate \(1-\alpha_L\) is supposed as large as possible. For a fixed LF model,  \(Q_{\tilde{\varepsilon}}\) increases monotonically with respect to both the sample size parameter \(n_L\) and the threshold parameter \(\tilde{\varepsilon}\). As \(n_L\) and \(\tilde{\varepsilon}\) increase, the algorithmic accuracy improves, while its efficiency decreases accordingly. Hence, \(n_L\) and \(\tilde{\varepsilon}\) represent a trade-off between the accuracy and efficiency. 
With a fixed LF model, adjusting the threshold parameter \(\tilde{\varepsilon}\) effectively controls the size of the region where \(Q_{\tilde{\varepsilon}} = 0\). Therefore, compared to tuning the sample size \(n_L\), adjusting \(\tilde{\varepsilon}\) plays a more critical role in influencing the algorithmic performance. In Section \ref{sec:maps}, we will embed the proposed method into a SMC framework, which allows adaptive selection of the threshold \(\tilde{\varepsilon}\), further improving algorithmic performance and robustness.

\subsection{Assessment of simulation model suitability}
Before employing the multifidelity approach, it is necessary to verify whether the selected or constructed LF model can serve as an effective pre-filtering simulator. In this section, we propose a metric that serves as a measure of consistency between the LF and HF simulation models. The basic idea of our metric is to evaluate the type II error rate (false negative ratio) of the pre-screening procedure, with fixed true positive rate. Here the true positive denotes accepted particles by the HF simulator that also pass the pre‑screening, and false negative denotes rejected particles by the HF simulator that pass the pre‑screening. This involves selecting a proper sets of thresholds (\(\varepsilon_0\), \(\tilde{\varepsilon}_0\)), and evaluating the false negative ratio. 

We first draw \(N_0\) (e.g., \(N_0 = 5000\)) parameter samples \(\{\boldsymbol{\theta}^{(i)}\}_{i=1}^{N_0}\) from the proposal distribution \(q(\cdot)\). For each parameter sample, we generate a set of LF simulation outputs \(\tilde{\boldsymbol{x}}^{(i)}_{1:n_L} \sim \tilde{p}(\cdot \mid \boldsymbol{\theta}^{(i)})\) and HF simulation outputs \(\boldsymbol{x}^{(i)}_{1:n_H} \sim p(\cdot \mid \boldsymbol{\theta}^{(i)})\).

{\bf Evaluating $\varepsilon_0$: }
We determine a HF distance threshold \(\varepsilon_0\) by solving \(PA(\{W_i\}_{i=1}^{N_0}, \varepsilon_0) = \kappa\) (e.g., \(\kappa = 0.1\)),  if $\varepsilon_0<\varepsilon$, set $\varepsilon_0 = \varepsilon$, where the weight for each sample is given by
\[
W_i \propto \frac{\pi(\boldsymbol{\theta}_i)}{q(\boldsymbol{\theta}_i)} \sum_{k=1}^{n_H} \mathbb{I}\left(\Delta(\boldsymbol{x}_{k}^{(i)}, \boldsymbol{y}) < \varepsilon_0\right),
\]
with \(\pi(\cdot)\) denoting the prior distribution and \(\mathbb{I}(\cdot)\) the indicator function. 
In practical applications, when \(\varepsilon\) is very small, the number of accepted samples after \(N_0\) trials may be insufficient, resulting in unreliable evaluation. 
This selection of $\varepsilon_0$ is to avoid the case that a too small $\varepsilon$ is used such that only  very few samples being retained.

{\bf Evaluating $\tilde{\varepsilon}_0$: }
The LF threshold \(\tilde{\varepsilon}_0\) is set as the maximum, over all samples with positive weights \(W_i > 0\), of their minimum LF distance to the observation:
\[
\tilde{\varepsilon}_0 = \max_{i: W_i > 0} \min_k \Delta(\tilde{\boldsymbol{x}}_{k}^{(i)}, \boldsymbol{y}).
\]
The intuition of selecting $\tilde{\varepsilon}_0$ is to guarantee that all the particles accepted by the HF model can pass the pre‑screening. Hence, the true positive rate is $1$ in this setting. 

{\bf Computing False Negative:}
For each parameter sample, we then count the number of LF simulations with a distance below \(\tilde{\varepsilon}_0\), denoted by
\[
\tilde{w}_i = \sum_{k=1}^{n_L} \mathbb{I}\left(\Delta(\tilde{\boldsymbol{x}}_{k}^{(i)}, \boldsymbol{y}) \le \tilde{\varepsilon}_0\right).
\]
Note that $\tilde{w}_i = 0$ only holds for particles rejected by both thresholds. 
Finally, we compute the metric
\[
E = \frac{1}{(1-\kappa) N_0} \sum_{i=1}^{N_0}\mathbb{I}(\tilde{w}_i > 0),
\]
which represents the false negative ratio.

This metric serves as a measure of consistency between the LF and HF simulation models. A lower value of $E$ indicates better consistency. 
In practical applications, the threshold \(\varepsilon_0\) is commonly set equal to \(\varepsilon\) to achieve optimal performance.

 \section{An adaptive pre-filtering multi-fidelity ABC-SMC algorithm}
\label{sec:maps}

 In this section, we embed the pre-filtering hierarchical importance sampling into an adaptive sequential Monte Carlo (SMC) framework and propose the Multifidelity approximate Bayesian computation with Adaptive Pre-filtering Sequential Monte Carlo (MAPS) algorithm. Building upon the framework of the standard SMC method with a decreasing threshold sequence \(\{\varepsilon_t\}\), we introduce a decreasing auxiliary threshold sequence \(\{\tilde{\varepsilon}_t\}\). These thresholds are alternately reduced to define an intermediate sequence of target distributions
\[
\pi_{\varepsilon_0}, \pi_{\varepsilon_0, \tilde{\varepsilon}_1}, \pi_{\varepsilon_1, \tilde{\varepsilon}_1}, \pi_{\varepsilon_1, \tilde{\varepsilon}_2}, \pi_{\varepsilon_2, \tilde{\varepsilon}_2}, \ldots,
\]
which progressively converges to 
\(\pi_{\varepsilon_T, \tilde{\varepsilon}_T}\), 
 where \(\varepsilon_T = \varepsilon\) is the final target threshold.

 Algorithm \ref{alg:MAPS-filter} provides an overview of the proposed MAPS algorithm. In the $t$-th SMC iteration, the algorithm iterates between \emph{determination of critical value}, \emph{auxiliary weighting}, \emph{selection of auxiliary thresholds}, \emph{resampling}, \emph{proposal}, \emph{weighting} and \emph{selection of thresholds} to achieve the approximation of $\pi_{\varepsilon_t, \tilde{\varepsilon}_t}$ from $\pi_{\varepsilon_{t-1}, \tilde{\varepsilon}_{t-1}}$. We detail these steps below. 

{\bf Determination of critical value: }
 The sequence of auxiliary thresholds \(\tilde{\varepsilon}_{1:T}\) affects the filtering performance and the final estimates of the posterior. To avoid selecting an overly conservative threshold (\eg~caused by outliers), we propose to determine a  lower bound for the auxiliary thresholds. 
 The critical value $\underline{\tilde{\varepsilon}}$ is selected by controlling \(a_L\), which represents the proportion of the LF simulations with distance falling above the critical value. 
 This \(a_L\) is typically set to a small positive value, such as 0.001. The numerical estimate of \(a_L\) is given by  
\[
a_L \approx \sum_{i=1}^N \tilde{w}_i \, \mathbb{I} \left( \min_k \Delta(\tilde{\xbold}_{t-1,k}^{(i)}, \ybold) \ge \underline{\tilde{\varepsilon}} \right),
\]
where \(\{(\boldsymbol{\theta}_t^{(i)}, \tilde{w}_i)\}_{i=1}^N\) denote weighted particles that approximate the ABC posterior distribution \(\pi_{\varepsilon_T}\).
This approximation of \(\pi_{\varepsilon_T}\) 
is achieved by an importance sampling procedure with an intermediate target as proposal distribution,  
and with weight update function
\[
\tilde{w}_i \propto W_{t-1}^{(i)} \frac{\sum_k \mathbb{I}\big(\Delta(\xbold_{t-1,k}^{(i)}, \ybold) < \varepsilon_T \big)}{\sum_k \mathbb{I}\big(\Delta(\xbold_{t-1,k}^{(i)}, \ybold) < \varepsilon_t \big)},
\]
where \(W_{t-1}^{(i)}\) represents the original weight of the \(i\)-th particle at iteration \(t\).
In practice, the critical value \(\underline{\tilde{\varepsilon}}\) is chosen as the \((1 - a_L)\)-quantile of the weighted minimum distances, i.e.,  
\[
\underline{\tilde{\varepsilon}} = Q_{1 - a_L} \left( \left\{ \min_k \Delta(\tilde{\xbold}_{t-1,k}^{(i)}, \ybold), \tilde{w}_i \right\} \right).
\]
As \(a_L\) approaches zero, the estimation error of the algorithm decreases. However, in practical applications, a strictly positive \(a_L\) is preferred over zero, since selecting the threshold as a quantile rather than the maximum value effectively controls estimation error while mitigating the influence of outliers. This approach prevents an excessively conservative threshold and consequently enhances the efficiency of the pre-filtering step.
In each iteration, the threshold \(\underline{\tilde{\varepsilon}}\) is updated because, as the algorithm progresses, the quality of the sample set \(\{(\boldsymbol{\theta}_t^{(i)}, \tilde{w}_i)\}_{i=1}^N\) improves, resulting in \(\underline{\tilde{\varepsilon}}\) converging increasingly closer to the true value.

 {\bf Update of auxiliary weights and thresholds:}
 We then reduce the auxiliary threshold \(\tilde{\varepsilon}\). The new auxiliary threshold \(\tilde{\varepsilon}_t\) is determined by retaining a fixed proportion \(\alpha\) of active particles from the previous weighted particle set \(\left\{(\thetabold_{t-1}^{(i)}, W_{t-1}^{(i)})\right\}_{i=1}^N\). Specifically, \(\tilde{\varepsilon}_t\) is chosen such that
\begin{equation}
    \operatorname{PA}\left(\{\tilde{W}_t^{(i)}\}; \tilde{\varepsilon}_t\right) = \alpha_L \operatorname{PA}\left(\{W_{t-1}^{(i)}\}\right),
\end{equation}
where the updated weights are defined by
\begin{equation}
    \tilde{W}_t^{(i)} \propto W_{t-1}^{(i)} \cdot \mathbb{I}\left\{\min_k \Delta(\xbold_{t-1,k}^{(i)}, \ybold) < \tilde{\varepsilon}_t \right\}.
\end{equation}
This yields a weighted particle set \(\left\{(\thetabold_{t-1}^{(i)}, \tilde{W}_t^{(i)})\right\}_{i=1}^N\), which approximately follows the distribution \({\pi}_{\varepsilon_{t-1}, \tilde{\varepsilon}_t}\). The reduction of the auxiliary threshold removes approximately \(1 - \alpha_L\) fraction of active particles. 
Finally, we set \(\tilde{\varepsilon}_t = \max\{\tilde{\varepsilon}_t, \underline{\tilde{\varepsilon}}\}\).
This mechanism ensures efficient screening while preventing the collapse of the algorithm due to an insufficient number of active particles.

{\bf Adaptive resampling:} 
To alleviate particle degeneracy, dynamic resampling is performed when the effective sample size (ESS) falls below a threshold $N_T$:
$$
\text{ESS} = {1}/{\sum_{i=1}^N (W_t^{(i)})^2} < N_T.
$$
After the resampling step, all particles are assigned equal weights.

{\bf Proposal:} We construct a Markov kernel with stationary distribution $\pi_{\varepsilon_{t-1},\tilde{\varepsilon}_{t}}$ to propagate these weighted particles, to enhance sample diversity. At each iteration, given the current state $\boldsymbol{z}_{t-1}=(\thetabold_{t-1}^{(i)}, \xbold_{t-1,1:n_H}^{(i)}, \tilde{\xbold}_{t-1,1:n_L}^{(i)})$, a new parameter $\thetabold^*$ is proposed according to the proposal distribution $q(\cdot \mid \thetabold_{t-1}^{(i)})$.  The MH acceptance probability for transitioning from the current state $\boldsymbol{z}_{t-1}$ to the proposed state $\boldsymbol{z}^{(i)*}=(\thetabold^{(i)*}, \xbold_{1:n_H}^{(i)*}, \tilde{\xbold}_{1:n_L}^{(i)*})$ is given by
$$
\tilde{\alpha}\left\{\boldsymbol{z}_{t-1}^{(i)},\boldsymbol{z}^{(i)*}\right\}
=
\frac{
\pi(\thetabold^{(i)*}) \displaystyle\sum_{k=1}^{n_H} \mathbb{I}\left\{\Delta(\xbold_k^{(i)*}, \ybold) < \varepsilon_{t-1}\right\} q(\thetabold_{t-1}^{(i)} \mid \thetabold^{(i)*}) \mathbb{I}\left\{\min_k \Delta(\tilde\xbold_{k}^{(i)*}, \ybold) < \tilde{\varepsilon}_{t}\right\}
}{
\pi(\thetabold_{t-1}^{(i)}) \displaystyle\sum_{k=1}^{n_H} \mathbb{I}\left\{\Delta(\xbold_k^{(i)*}, \ybold) < \varepsilon_{t-1}\right\} q(\thetabold^{(i)*} \mid \thetabold_{t-1}^{(i)}) \mathbb{I}\left\{\min_k \Delta(\tilde\xbold_{k}^{(i)*}, \ybold) < \tilde{\varepsilon}_{t}\right\}
},
$$
where HF simulations $\xbold_{1:n_H}^{(i)*} \sim (\cdot \mid \thetabold^{(i)*})$, and LF simulations $\tilde{\xbold}_{1:n_L}^{(i)*} \sim \tilde{p}(\cdot\mid\thetabold^{(i)*})$.
Since the current state is already drawn from $\pi_{\varepsilon_{t-1},\tilde{\varepsilon}_t}$, the factor $\mathbb{I}\left\{\min_k \Delta(\tilde\xbold_{k}, \ybold)< \tilde{\varepsilon}\right\}$ in the denominator is always one. Thus, the MH acceptance probability simplifies to
\begin{equation}
\label{eq: PF-MH-acc}
\tilde{\alpha}\left\{\boldsymbol{z}_{t-1}^{(i)},\boldsymbol{z}^{(i)*}\right\}
=
\frac{
\pi(\thetabold^{(i)*}) \displaystyle\sum_{k=1}^{n_H} \mathbb{I}\left\{\Delta(\xbold_k^{(i)*}, \ybold) < \varepsilon_{t-1}\right\} q(\thetabold_{t-1}^{(i)} \mid \thetabold^{(i)*}) \mathbb{I}\left\{\min_k \Delta(\tilde\xbold_{k}^{(i)*}, \ybold) < \tilde{\varepsilon}_{t}\right\}
}{
\pi(\thetabold_{t-1}^{(i)}) \displaystyle\sum_{k=1}^{n_H} \mathbb{I}\left\{\Delta(\xbold_k^{(i)*}, \ybold) < \varepsilon_{t-1}\right\} q(\thetabold^{(i)*} \mid \thetabold_{t-1}^{(i)}) 
}.
\end{equation}
It is evident from Eq.\eqref{eq: PF-MH-acc} that when \(\min_k \Delta(\tilde\xbold_{k}^{(i)*}, \ybold) \geq \tilde{\varepsilon}_t\), the acceptance probability is zero. Therefore, one can first perform LF simulations for the proposed parameter \(\thetabold^{(i)*}\). Only if the LF discrepancy satisfies \(\min_k \Delta(\tilde\xbold_{k}^{(i)*}, \ybold) < \tilde{\varepsilon}_t\) should HF simulations be carried out. In this case, the candidate parameter is accepted with the Metropolis–Hastings acceptance probability given by Eq.\eqref{eq: PF-MH-acc}. Otherwise, the proposed parameter can be immediately rejected without performing HF simulations. The detailed procedure is summarized in Algorithm \ref{alg:PF-MCMC}.
\begin{algorithm}[H]
\caption{Single stage of pre-filtering ABC-MCMC}
\label{alg:PF-MCMC}
{\bf Procedure} PF-ABC-MCMC($(\thetabold_{t-1}^{(i)},\xbold_{t-1,1:n_H}^{(i)},\tilde{\xbold}_{t-1,1:n_L}^{(i)})$, $\pi$, $q$, $\tilde{P}$, $P$, $\tilde{\varepsilon}$, $\varepsilon$)

{\bf Input: } Prior: $\pi(\cdot)$, LF simulator $\tilde{P}_\thetabold$,  HF simulator $P_\thetabold$, observation data $\ybold$, threshold $\varepsilon_{t-1}$ and $\tilde{\varepsilon}_t$, current state $(\thetabold_{t-1}^{(i)},\xbold_{t-1,1:n_H}^{(i)},\tilde{\xbold}_{t-1,1:n_L}^{(i)})$.

{\bf Output:} $(\thetabold_{t}^{(i)},\xbold_{t,1:n_H}^{(i)},\tilde{\xbold}_{t,1:n_L}^{(i)})$

\begin{algorithmic}[1]
\STATE Sample $\thetabold^{(i)*}\sim q(\cdot\mid\thetabold_{t-1}^{(i)})$ and $\tilde{\xbold}^{(i)*}_{k}\sim \tilde{P}_{\thetabold^{(i)*}}$, where $k=1,\ldots,n_L$. 
\IF{$\min_k \Delta(\tilde\xbold_{k}^{(i)*}, \ybold)>\tilde{\varepsilon}_t$}
\STATE Set $\left\{\thetabold_t^{(i)},\xbold_{t,1:n_H}^{(i)},\tilde{\xbold}_{t,1:n_L}^{(i)}\right\}=\left\{\thetabold_{t-1}^{(i)},\xbold_{t-1,1:n_H}^{(i)},\tilde{\xbold}_{t-1,1:n_L}^{(i)}\right\}$.
\ELSE 
\STATE Sample $\xbold_{t,1:n_H}^{(i)*} \sim P_{\thetabold^{(i)*}}$, set $\left\{\thetabold_{t}^{(i)},\xbold_{t,1:n_H}^{(i)},\tilde{\xbold}_{t,1:n_L}^{(i)}\right\}=\left\{\thetabold^{(i)*},\xbold_{1:n_H}^{(i)*},\tilde{\xbold}_{1:n_L}^{(i)*}\right\}$ with probability Eq.\eqref{eq: PF-MH-acc}; otherwise set $\left\{\thetabold_t^{(i)},\xbold_{t,1:n_H}^{(i)},\tilde{\xbold}_{t,1:n_L}^{(i)}\right\}=\left\{\thetabold_{t-1}^{(i)},\xbold_{t-1,1:n_H}^{(i)},\tilde{\xbold}_{t-1,1:n_L}^{(i)}\right\}$.
\ENDIF

\end{algorithmic}
\end{algorithm}

 {\bf Update of weights and thresholds:}
We determine the threshold $\varepsilon_t$ by making  
$\operatorname{PA}\left(\left\{W_t^{(i)}\right\};\varepsilon_t\right)=\alpha\operatorname{PA}\left(\left\{\tilde{W}_{t}^{(i)}\right\}\right)$, where \begin{equation}
    W_t^{(i)} \propto \tilde{W}_{t}^{(i)} \frac{\sum_k\left\{\mathbb{I}(\Delta({\xbold}_{t-1,k}^{(i)},\ybold)<\varepsilon_t\right\}}{\sum_k\left\{\mathbb{I}(\Delta({\xbold}_{t-1,k}^{(i)},\ybold)<\varepsilon_{t-1}\right\}}.
\end{equation}
The Markov kernel transition is carried out before determining the new threshold \(\varepsilon_t\) because the auxiliary threshold \(\tilde{\varepsilon}_t\) is obtained by tightening the condition relative to the previous threshold \(\varepsilon_{t-1}\). Given that \(\tilde{\varepsilon}_t\) is more stringent, the Markov kernel transition targeting the distribution \(\pi_{\varepsilon_{t-1}, \tilde{\varepsilon}_t}\) benefits more significantly from the pre-filtering effect of the LF simulations. This leads to a more efficient particle refinement and accelerates convergence within the SMC framework.

\begin{figure}[H]
    \centering
    \includegraphics[width=1\linewidth]{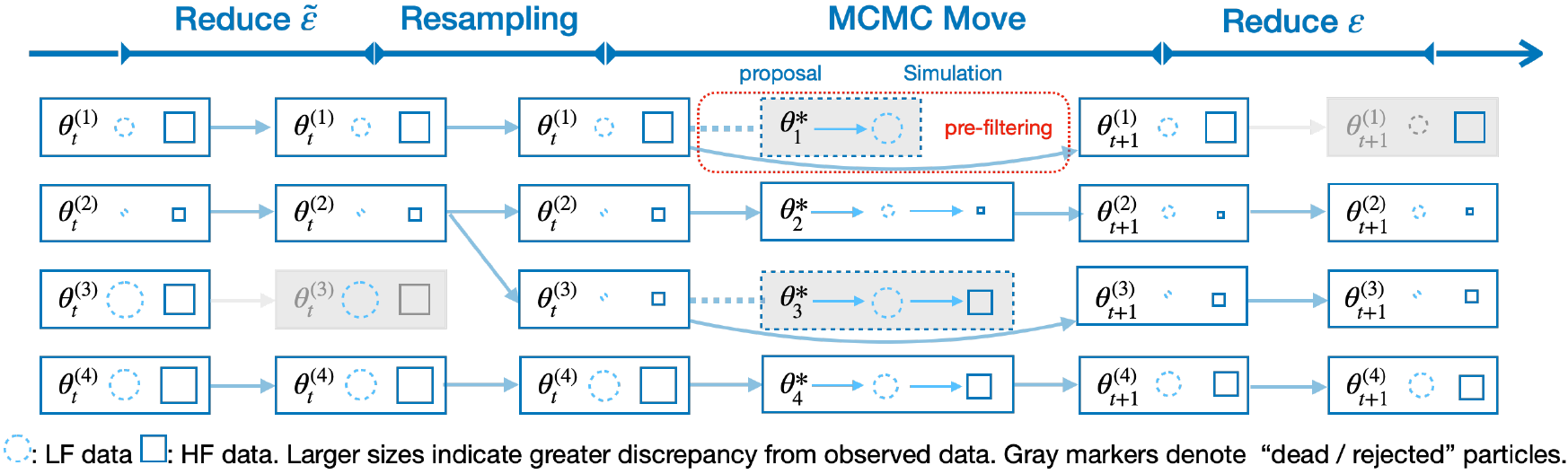}
    \caption{An overview of the MAPS algorithm (Steps 6-10 of Algorithm \ref{alg:MAPS-filter}). Given a set of weighted particles $\thetabold_t^{1:N}$, the MAPS algorithm performs the following steps to approximate $\thetabold_{t+1}^{1:N}$: (1)~decreasing the auxiliary threshold to filter out low-activity particles; 
    (2)~adaptive resampling based on effective sample size; 
    (3)~multi-fidelity MCMC moves with LF pre-filtering and HF refinement; 
    (4)~final threshold reduction.}
    \label{fig:MAPS-flow chart}
\end{figure}

In practice, the HF simulation in Algorithm \ref{alg:MAPS-filter} Step~\ref{alg:MAPS-filter, gen} can be omitted. This is because, at $t=1$ with $\varepsilon_0 = \infty$, the Metropolis-Hastings acceptance probability in Step~\ref{alg:MAPS-filter, MCMC} of the algorithm is given by
\[
\tilde{\alpha}\{\mathbf{z}, \mathbf{z}^*\} =
\frac{
\pi(\boldsymbol{\theta}^*)q(\boldsymbol{\theta} \mid \boldsymbol{\theta}^*) \mathbb{I}\left\{\min_k \Delta(\tilde{\xbold}_{k}^{*}, \ybold) < \tilde{\varepsilon}_{t}\right\}
}{
\pi(\boldsymbol{\theta}) q(\boldsymbol{\theta}^* \mid \boldsymbol{\theta})
}.
\]
At this stage, HF simulated data is not required. Instead, it is sufficient to conduct HF simulations only for the accepted samples after the first MCMC move, so that these HF data can be used in subsequent steps. Figure \ref{fig:MAPS-flow chart} presents a flowchart of the MAPS algorithm (corresponding to Steps 6–10 in Algorithm \ref{alg:MAPS-filter}), which provides a more intuitive understanding of the algorithmic procedure.

\begin{algorithm}[H]
\caption{MAPS: Multifidelity ABC with pre-filtering SMC}
\label{alg:MAPS-filter}
{\bf Input: } Prior: $\pi(\cdot)$, LF simulator $\tilde{p}(\cdot\mid\cdot)$,  HF simulator $p(\cdot\mid\cdot)$, observation data $\ybold$, target threshold $\varepsilon_T$, number of particles at each iteration $N$, ESS threshold $N_T$, preserved proportion of active particles $\alpha$(based on HF data) and $\alpha_L$ (based on LF data).  

{\bf Output:} $\left\{\thetabold_t^{(i)},W_t^{(i)}\right\}_{i=1:N}$.
\begin{algorithmic}[1]
\STATE Initialize $\varepsilon_0=\infty$, $\tilde\varepsilon_0=\infty$, $t=0$. 
\FOR{$i=1,\ldots, N$}
\STATE Sample $\boldsymbol{\theta}_0^{(i)} \sim \pi(\cdot)$ and set $W_0^{(i)}=1/N$. \textcolor{grey}{ \%$\{(\tilde{W}_0^{(i)},\thetabold_{0}^{(i)})\}_{i=1:N}\sim \pi_{\varepsilon_{0}}$.} 
\STATE  Generate $\tilde{\xbold}^{(i)}_{0,1:n_L}\sim \tilde{p}(\cdot\mid\thetabold_0^{(i)})$ and ${\xbold}^{(i)}_{0,1:n_H}\sim {p}(\cdot\mid\thetabold_0^{(i)})$. 
\label{alg:MAPS-filter, gen}
\ENDFOR

\WHILE{$\varepsilon_{t}>\varepsilon_T$}

\STATE Set $\tilde{w}_i\propto {W}_{t}^{(i)}\frac{\sum_k\mathbb{I}(\Delta({\xbold}_{t,k}^{(i)},\ybold)<\varepsilon_T)}{\sum_k\mathbb{I}(\Delta({\xbold}_{t,k}^{(i)},\ybold)<\varepsilon_{t})},$ $\underline{\tilde{\varepsilon}}= Q_{1-a_L}(\left\{\min_k\Delta(\tilde{\xbold}_{i,k}^{(i)},\ybold),\tilde{w}_i\right\}) $.\label{alg:MAPS-filter, Aeps}
\STATE Set $t=t+1$,  determine $\tilde{\varepsilon}_t$ by making  $\operatorname{PA}\left(\left\{\tilde{W}_t^{(i)}\right\},\tilde{\varepsilon}_t\right)=\alpha_L\operatorname{PA}\left(\left\{W_{t-1}^{(i)}\right\}\right)$, if $\tilde{\varepsilon}_t<\underline{\tilde{\varepsilon}}$, set $\tilde{\varepsilon}_t=\tilde{\varepsilon}_T$, where $\tilde{W}_t^{(i)} \propto W_{t-1}^{(i)} {\mathbb{I}\left\{\min_k\Delta(\tilde{\xbold}_{t-1,k}^{(i)},\ybold)<\tilde{\varepsilon}_t\right\}}$.  \textcolor{grey}{ \%$\{(\tilde{W}_t^{(i)},\thetabold_{t-1}^{(i)})\}_{i=1:N}\sim \pi_{\varepsilon_{t-1},\tilde{\varepsilon}_t}$.}\label{alg:MAPS-filter, MCMC}
\IF{$\text{ESS}\left\{\tilde{W}_t^{(i)}\right\}<N_T$}
\STATE Resample $N$ particles from 
$\left\{\thetabold_{t-1}^{(i)},\xbold_{t-1,1:n_H}^{(i)},\tilde{\xbold}_{t-1,1:n_L}^{(i)}\right\}$ with weight $\left\{\tilde{W}_t^{(i)}\right\}$, also denoted abusively $\left\{\thetabold_{t-1}^{(i)},\xbold_{t-1,1:n_H}^{(i)},\tilde{\xbold}_{t-1,1:n_L}^{(i)}\right\}$ and set $\tilde{W}_t^{(i)}=\frac{1}{N}$.
\ELSE
\STATE Set $\left\{\thetabold_{t-1}^{(i)},\xbold_{t-1,1:n_H}^{(i)},\tilde{\xbold}_{t-1,1:n_L}^{(i)}\right\} = \left\{\thetabold_{t-1}^{(i)},\xbold_{t-1,1:n_H}^{(i)},\tilde{\xbold}_{t-1,1:n_L}^{(i)}\right\}$, and $\left\{\tilde{W}_t^{(i)}\right\} = \left\{\tilde{W}_t^{(i)}\right\}$
\ENDIF
\FOR{$i=1,\ldots, N$}
\STATE Obtain $\left\{\thetabold_{t}^{(i)},\xbold_{t,1:n_H}^{(i)},\tilde{\xbold}_{t,1:n_L}^{(i)}\right\}$ by running PF-ABC-MCMC (Algorithm~\ref{alg:PF-MCMC}). 
\ENDFOR
\STATE Determine $\varepsilon_t$ by making  $\operatorname{PA}\left(\left\{W_t^{(i)}\right\},\varepsilon_t\right)=\alpha\operatorname{PA}\left(\left\{\tilde{W}_{t}^{(i)}\right\}\right)$, and if $\varepsilon_t<\varepsilon_T$, set $\varepsilon_t=\varepsilon_T$, where $$W_t^{(i)} \propto \tilde{W}_{t}^{(i)}\frac{\sum_k\mathbb{I}(\Delta({\xbold}_{t,k}^{(i)},\ybold)<\varepsilon_t)}{\sum_k\mathbb{I}(\Delta({\xbold}_{t,k}^{(i)},\ybold)<\varepsilon_{t-1})}.$$ \textcolor{grey}{ \%$\{(\tilde{W}_t^{(i)},\thetabold_{t}^{(i)})\}_{i=1:N}\sim \pi_{\varepsilon_{t},\tilde{\varepsilon}_t}$.}
\ENDWHILE
\end{algorithmic}
\end{algorithm}

\section{Numerical experiments}
\label{sec:numerical}

\subsection{Toy example}
\label{sec:toy}
In this section, we utilize a toy example to demonstrate the advantages of our proposed MAPS algorithm compared to the standard ABC-ASMC algorithm. The HF simulation model is defined as \(x \mid \theta \sim N(4\theta^2 + 0.3\cos(5\pi\theta), 0.2^2)\), and the LF simulation model is defined as \(x \mid \theta \sim N(4\theta^2, 0.2^2)\). We set the prior distribution as \(\mathcal{U}[-2, 2]\) and the target threshold \(\varepsilon_T = 0.1\).

\begin{figure}[H]
    \centering
    \begin{subfigure}[b]{0.45\linewidth}
        \includegraphics[width=\linewidth]{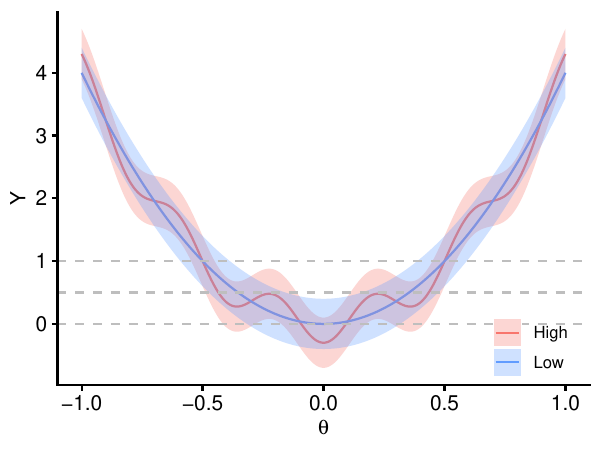}
        \caption{}
        \label{fig:points}
    \end{subfigure}
    \hfill
    \begin{subfigure}[b]{0.45\linewidth}
        \includegraphics[width=\linewidth]{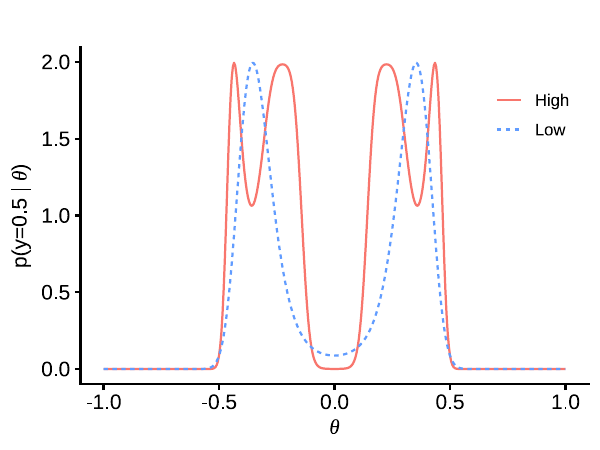}
        \caption{}
        \label{fig:pyobs05}
    \end{subfigure}
    \caption{(a) Variation of HF and LF simulation data as a function of the parameter \(\theta\). Solid lines denote mean values; shaded areas represent 95\% confidence intervals. Gray dashed lines mark observed values \(y_{\text{obs}}=0\), \(0.5\), and \(1\). (b) Likelihood functions for both fidelity models at \(y_{\text{obs}}=0.5\).}
    \label{fig:toyexample}
\end{figure}

Figure \ref{fig:toyexample}(\subref{fig:points}) illustrates how the LF and HF simulation data vary with the parameter \(\theta\). Overall, the LF simulation serves as an approximation of the HF simulation, though notable differences exist in certain regions. We consider three sets of observational data: \(y_{\text{obs}} = 0\) (where the HF likelihood is multimodal while the LF likelihood is unimodal), \(y_{\text{obs}} = 0.5\) (where the maximum points of the HF and LF likelihoods differ), and \(y_{\text{obs}} = 1\) (which reveals regions where the HF likelihood is significantly greater than zero, whereas the LF likelihood approaches zero).  In the main text, we present only the results for \(y_{\text{obs}} = 0.5\), the corresponding likelihood functions are shown in Figure \ref{fig:toyexample} (\subref{fig:pyobs05}), with the remaining results provided in Appendix Section S.2.1.

The discrepancy function is defined as \(\Delta(x, y) = (x - y)^2\), and the number of particles in each iteration is set to \(N = 5120\). In the standard ABC-ASMC algorithm, the number of simulations for each parameter is \(n = 10\), with a threshold reduction parameter \(\alpha = 0.7\). In the MAPS algorithm, the number of simulations for each parameter is \(n_H = 10,~n_L = 20\) with threshold reduction parameter \(\alpha = 0.7\). We investigated the impact of the auxiliary threshold reduction parameter $\alpha_L$ and the auxiliary critical value parameter $a_L$ on both algorithmic efficiency and result accuracy. The results are presented in Figures S.1 and S.2 of Appendix Section S.2.1.  Based on the numerical results, we recommend to set $\alpha_L=\alpha=0.7$ and $a_L=0.001$.

\begin{figure}[H]
    \centering
    \includegraphics[width=1\linewidth]{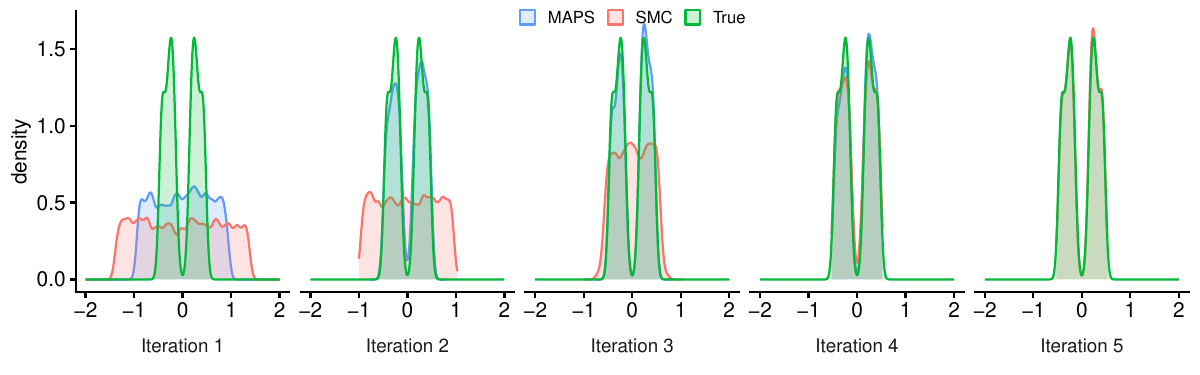}
    \caption{Comparison of posterior density estimates over iterations for ABC-ASMC and MAPS methods, as well as the true posterior density, when $y_{obs} = 0.5$.}
    \label{fig:toy-ite-density0.5}
\end{figure}

\begin{figure}[H]
    \centering    \includegraphics[width=1\linewidth]{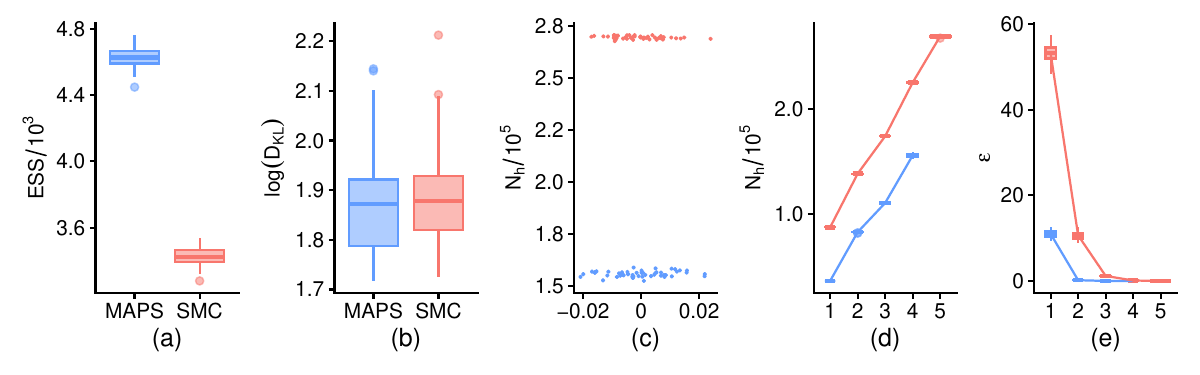}
    \caption{Comparison of results from 50 repetitions of ABC-ASMC and MAPS methods with $y_{\mathrm{obs}} = 0.5$. (a) effective sample size, (b) KL divergence, (c) posterior mean and total number of HF simulations, (d) the threshold evolution and (e) the number of HF simulations as a function of iterations.}
    \label{fig:toy_example_summary0.5}
\end{figure}

Figure \ref{fig:toy-ite-density0.5} presents the posterior density estimates obtained by ABC-ASMC and MAPS across iterations, alongside the true density estimate. As shown in the figure, by incorporating LF simulations for pre-filtering, the MAPS method achieves a faster convergence of the posterior density to the true ABC posterior compared to ABC-ASMC.  
Figure \ref{fig:toy_example_summary0.5} presents a comparison of summary of the results from 50 repeated runs of MAPS and ABC-ASMC, where the true posterior is obtained via analytical calculation. The results demonstrate that the MAPS method outperforms ABC-ASMC in terms of effective sample size and KL divergence. In addition, the number of HF simulations required by MAPS is significantly lower than that of ABC-ASMC. The middle panel shows the relationship between the estimated posterior mean and the number of simulations for both MAPS and ABC-ASMC, further highlighting that MAPS achieves comparable or even superior inference accuracy while substantially reducing the number of HF simulations. The last two panels of Figure \ref{fig:toy_example_summary0.5} illustrate the evolution of the threshold \(\varepsilon\) and the number of HF simulations across iterations. Due to the introduction of pre-filtering, the  threshold \(\varepsilon\) of MAPS converges more rapidly to the target value; specifically, ABC-ASMC requires five iterations, whereas MAPS converges in four.

Table \ref{tab:toy} summarizes the comparison between MAPS and ABC-ASMC methods for three different observed data points \( y_{\text{obs}} = 1 \), \( y_{\text{obs}} = 0.5 \), and \( y_{\text{obs}} = 0 \). The evaluation metrics include the Kullback–Leibler (KL) divergence, effective sample size, and the total number of HF simulations. All reported values represent averages over 50 independent repetitions. Percentage changes indicate the relative improvement of MAPS over ABC-ASMC. 
The results demonstrate that MAPS consistently achieves a substantial reduction in the number of HF simulations across all tested scenarios. Specifically, MAPS reduces HF simulations by approximately 34\% to 42\% compared to ABC-ASMC, reflecting a significant computational efficiency gain. In terms of inference accuracy, as measured by KL divergence, MAPS yields comparable or improved performance, with improvements up to 21\% for \( y_{\text{obs}}=0.5 \). The effective sample size of MAPS is also generally higher, indicating more reliable posterior estimates, with increments exceeding 30\% in two cases. Notably, for the case \( y_{\text{obs}}=0 \), while the KL divergence and effective sample size are similar or slightly worse, the method still achieves a marked reduction in computational cost.
Overall, these results confirm that MAPS provides a more efficient inference framework, significantly reducing the reliance on expensive HF simulations without compromising, and in some cases even enhancing, the quality of posterior estimation.
\begin{table}[H]
    \centering
    \caption{Comparison of KL divergence, effective sample size, and total number of HF simulations between SMC and MAPS methods for different observed values $y_{{obs}}$. Values are averages over 50 repetitions. Percentage changes indicate relative improvement of MAPS over SMC.}
    \begin{tabular}{c|cc cc|cccc|cccc}
    \hline
    &\multicolumn{4}{c}{KL divergence}&\multicolumn{4}{|c}{Effective sample size}&\multicolumn{4}{|c}{Number of HF simulations}\\    \hline
    &  SMC& &\multicolumn{2}{c|}{MAPS}&  SMC &&\multicolumn{2}{c|}{MAPS}&  SMC& &\multicolumn{2}{c}{MAPS} \\
        $y_{obs}=1$ &  0.04&&0.039& \textbf{-2.4\% }
        & 1157&&1614&\textbf{39.4\%}
        & 327652&&196979 & \textbf{-39.9\%}\\
        $y_{obs}=0.5$ & 0.071&&0.056&\textbf{-21.1}\%
        & 3426 &&4628   &     \textbf{35.1}\%
        & 269479  &&155677 &   \textbf{-42.2}\%\\
        $y_{obs}=0$&0.152&& 0.153&0.1\%
        & 3816&&3621&    -5.1\%
        & 319667&&210058&   \textbf{-34.3}\%\\\hline
    \end{tabular}
    \label{tab:toy}
\end{table}

\subsection{Ornstein-Uhlenbeck process}
\label{sec:OU}
In this subsection, we consider the Ornstein-Uhlenbeck process defined by
\[
dX_t = \gamma(\mu - X_t) \, dt + \sigma \, dW_t,
\]
where the initial condition is distributed as \(X(0) \sim \mathcal{N}(\mu + \mu_{\mathrm{offset}}, 0.1^2)\). Here, \(\mu\) denotes the long-term mean, \(\sigma\) is the diffusion coefficient, \(\gamma\) represents the rate of mean reversion, and \(\mu_{\mathrm{offset}}\) quantifies the deviation of the initial state from \(\mu\). The HF simulation is approximated using the Euler–Maruyama scheme:
\[
X(t + \delta_t) = X(t) + \gamma(\mu - X_t) \, \delta_t + \sigma \sqrt{\delta_t} \, z,
\]
where $z\sim \mathcal{N}(0,1)$.
The LF model assumes a stationary distribution given by
\[
X_t \sim \mathcal{N} \left( \mu, \left(\frac{\sigma}{2.5\gamma}\right)^2 \right).
\]
The prior distributions for the parameters are specified as follows:
\[
\mu \sim \mathcal{U}(0.1, 3), \quad \sigma \sim \mathcal{U}(0.1, 1), \quad \gamma \sim \mathcal{U}(0.1, 2), \quad \mu_{\mathrm{offset}} \sim \mathcal{U}(2, 6).
\]
The HF data, denoted as \(\xbold\), are generated from the HF model for a given set of parameters using a time step of \(\delta_t = 0.01\) over the interval \([0, 30]\). Observations are collected by sampling the model output at intervals with length \(t = 0.1\), resulting in \(\xbold = (x_1, x_2, \ldots, x_{301})\), where each \(x_j\) corresponds to the model state at the respective time point. 
And observational data \(\ybold\) were generated by performing HF simulations with parameter values \(\mu = 2\), \(\sigma = 0.5\), \(\gamma = 1\), and \(\mu_{\mathrm{offset}} = 3\). 
We define the summary statistics \(\boldsymbol{S}(\xbold)\) of a trajectory \(\xbold\) as:
\[
S_1(\xbold) = \frac{1}{150} \sum_{j=151}^{301} x_j, \quad
S_2(\xbold) = 10 \times \mathrm{sd}(x_{151:301}), \quad
S_3(\xbold) = x_1 - S_1(\xbold), \quad
S_4(\xbold) = x_1 - x_{21},
\]
where \(\mathrm{sd}(\cdot)\) denotes the sample standard deviation. These four summary statistics each serve as a rough estimator for one of the four parameters of interest. The second statistic includes a multiplicative factor of 10 to ensure that its scale is comparable to the other summary statistics.
The discrepancy measure between the HF simulated data \(\xbold\) and observational data \(\ybold\) is defined as
\[
\Delta(\xbold, \ybold) = \frac{1}{4} \left\| \boldsymbol{S}(\xbold) - \boldsymbol{S}(\ybold) \right\|_2^2.
\]

\begin{figure}[H]
    \centering
    \includegraphics[width=0.5\linewidth]{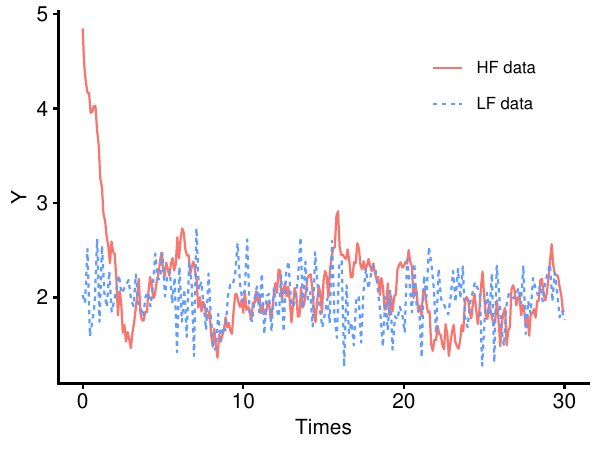}
    \caption{HF and LF simulation data generated using parameters \(\mu = 2\), \(\sigma = 0.5\), \(\gamma = 1\), and \(\mu_{\mathrm{offset}} = 3\).}
    \label{fig:OU_yobs}
\end{figure}

LF data \(\tilde{\xbold}\) are generated from the LF model for a given parameter set, consisting of 200 simulated points. The corresponding summary statistics \(\tilde{\boldsymbol{S}}(\tilde{\xbold})\) are defined as
\[
S_1(\tilde{\xbold}) = \frac{1}{200} \sum_{i=1}^{200} \tilde{x}_i, \quad
S_2(\tilde{\xbold}) = 10 \times \mathrm{sd}(\tilde{\xbold}).
\] The discrepancy between LF simulated data and observational data is measured by
\[
\Delta(\tilde{\xbold}, \ybold) = \frac{1}{2} \left\| \tilde{\boldsymbol{S}}(\tilde{\xbold}) - \boldsymbol{S}_{1,2}(\ybold) \right\|_2^2,
\]
where \(\boldsymbol{S}_{1,2}(\ybold)\) denotes the first two components of \(\boldsymbol{S}(\ybold)\).
Figure \ref{fig:OU_yobs} presents the simulation data generated from the HF and LF models using the true parameter values.

\begin{figure}[H]
    \centering
    \includegraphics[width=1\linewidth]{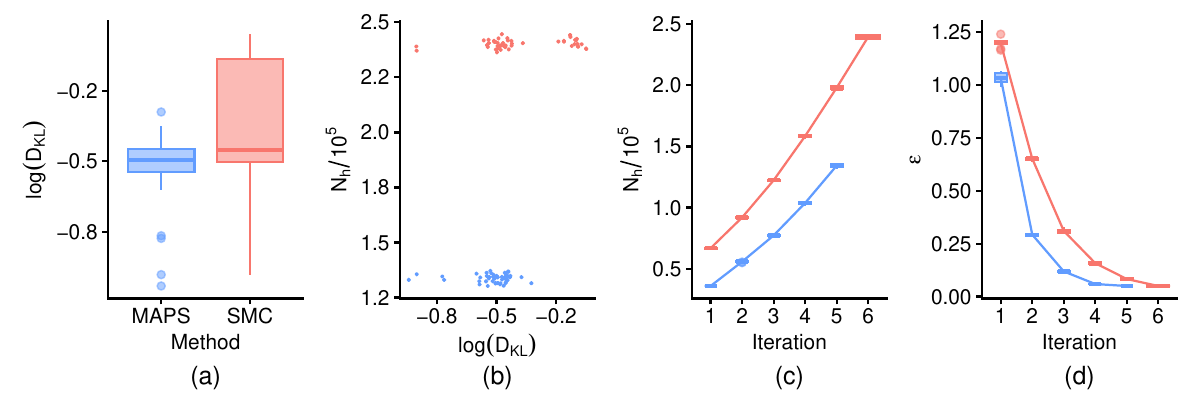}
    \caption{Comparison of results from 50 repetitions of ABC-ASMC and MAPS methods.  (a) KL divergence, (b) KL divergence and total number of HF simulations, (c) threshold evolution, and (d)number of HF simulations across iterations.}
    \label{fig:OU_summary}
\end{figure}

Figure \ref{fig:OU_summary} compares the performance of the MAPS and ABC-ASMC algorithms, each employing 5,120 particles and repeated 50 times to ensure statistical robustness. For the MAPS algorithm, the number of simulations per parameter is set to \(n_H = 10\) for the HF model and \(n_L = 20\) for the LF model, with threshold reduction parameters \(\alpha_H = \alpha_L = 0.7\) and a baseline threshold \(a_L = 0.001\). The standard ABC-ASMC algorithm uses \(n = n_H = 10\) simulations per parameter with a threshold reduction parameter \(\alpha = 0.7\). The true ABC posterior is approximated via ABC-MCMC using 10 chains, each with 500,000 iterations, where the first 100,000 iterations are discarded as burn-in.

\begin{figure}[H]
    \centering
    \begin{subfigure}[b]{0.32\linewidth}
        \centering
        \includegraphics[width=\linewidth]{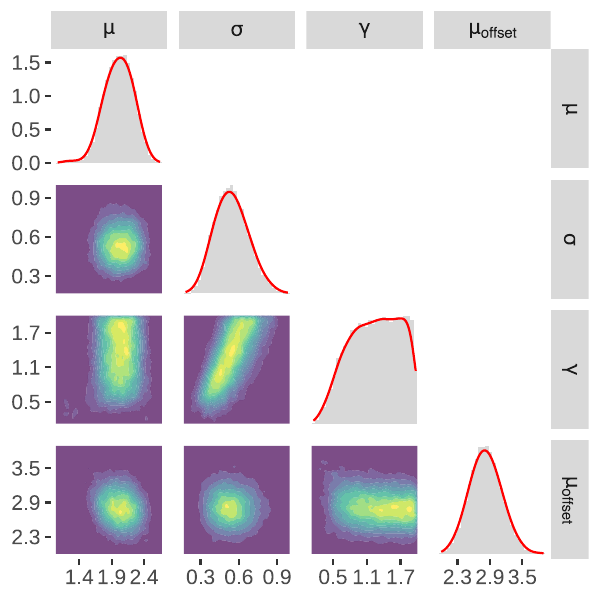}
        \caption{True posterior}
        \label{fig:OU_true}
    \end{subfigure}
    \hfill
    \begin{subfigure}[b]{0.32\linewidth}
        \centering
        \includegraphics[width=\linewidth]{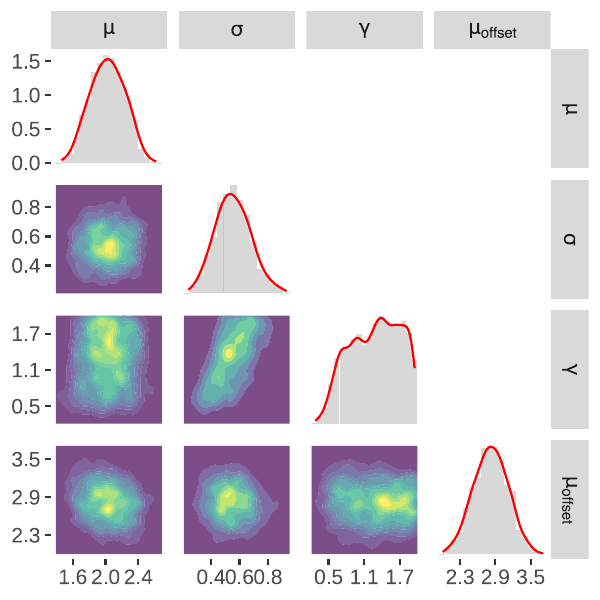}
        \caption{ABC-ASMC}
        \label{fig:OU_SMC}
    \end{subfigure}
    \hfill
    \begin{subfigure}[b]{0.32\linewidth}
        \centering
        \includegraphics[width=\linewidth]{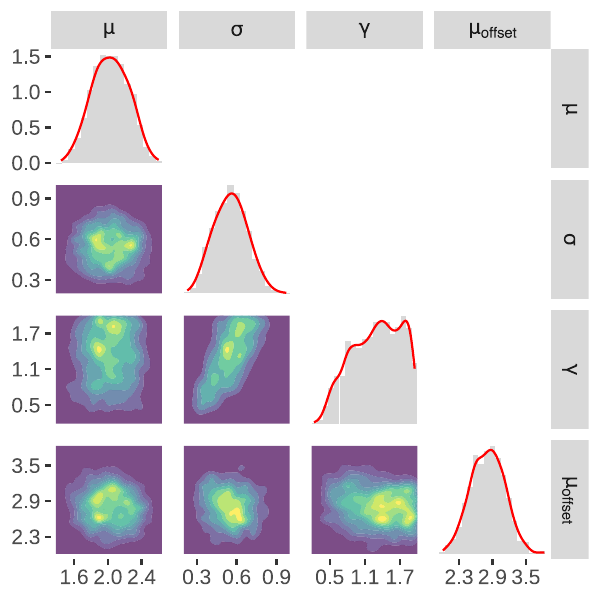}
        \caption{MAPS}
        \label{fig:OU_MAPS}
    \end{subfigure}
    \caption{Comparison of two-dimensional and one-dimensional marginal posterior distributions.}
    \label{fig:OU-2d}
\end{figure}

From the results depicted in Figure \ref{fig:OU_summary}, the MAPS algorithm achieves superior inference results with a significantly reduced number of HF simulations—approximately 44\% fewer than ABC-ASMC. This improvement is reflected in a consistently lower KL divergence relative to the true ABC posterior, indicating enhanced posterior approximation accuracy. Moreover, Figure \ref{fig:OU-2d} shows that the MAPS posterior closely matches both the true posterior and the ABC-ASMC estimate, confirming the effectiveness of the MAPS approach.

\subsection{Kuramoto oscillator network}
\label{sec:KU}
In this section, we consider the Kuramoto oscillator model as studied in \cite{prescott2021multifidelity}. The model consists of a fully connected network of \(M\) oscillators, each with phase \(\varphi_i(t)\) governed by
\begin{equation}
\label{eq:Ko-high}
    \dot{\varphi}_i = \omega_i + \frac{K}{M} \sum_{j=1}^M \sin(\varphi_j - \varphi_i), \quad i=1,\ldots,M,
\end{equation}
where \(\omega_i\) are independent draws from a \(\mathrm{Cauchy}(\omega_0, \gamma)\), and \(K\) represents the coupling strength. Initial phases are set to zero and simulations run over \(t \in [0,T]\).

The multifidelity approach leverages a low-dimensional representation of the coupled oscillator system by monitoring the Daido order parameters, which are defined as
\[
Z_n(t) = \frac{1}{M} \sum_{j=1}^{M} \exp(i n \varphi_j(t)),
\]
where \(n\) is a positive integer and \(i\) denotes the imaginary unit. In particular, the first-order parameter \(Z_1(t)\) is commonly expressed in terms of its magnitude \(R(t) = |Z_1(t)|\) and phase \(\Phi(t) = \arg(Z_1(t))\).

Under the Ott–Antonsen ansatz, the relationship \(Z_n(t) = [Z_1(t)]^n\) holds, allowing the system to be reduced to a single ordinary differential equation (ODE) governing \(Z_1\). Consequently, the original \(M\)-dimensional ODE system described in Equation \eqref{eq:Ko-high} can be effectively approximated by the following two-dimensional ODE system:
\begin{equation}
\label{eq:Ko-low}
    \dot{\tilde{R}} = \left(\frac{K}{2} - \gamma\right)\tilde{R} - \frac{K}{2} \tilde{R}^3, \quad \dot{\tilde{\Phi}} = \omega_0,
\end{equation}
with initial conditions \(\tilde{R}(0) = 1\) and \(\tilde{\Phi}(0) = 0\). This reduction facilitates efficient computation while preserving the essential dynamics of the oscillator ensemble.

Our goal is to infer parameters \((K, \omega_0, \gamma)\) from synthetic data generated by simulating \(M=32\) oscillators with \(\omega_i \sim \mathrm{Cauchy}(\omega_0, \gamma)\) over \(t \in (0,20]\). 
To summarize the trajectories for comparison, three statistics are used:
\[
S_1 = \left(\frac{1}{20} \int_0^{20} R(t) dt \right)^2, \quad S_2 = \frac{\Phi(20) - \Phi(0)}{20}, \quad S_3 = R(T_{1/2}),
\]
where \(T_{1/2}\) is the earliest time at which \(R_{\mathrm{obs}}\) reaches the midpoint between its initial and long-term average values. 

The HF simulation data \((R, \Phi)\) were generated from Eq.~\eqref{eq:Ko-high} using the classical fourth-order Runge-Kutta method with a time step of $0.1$. The LF simulation data \((\tilde{R}, \tilde{\Phi})\) were obtained from Eq.~\eqref{eq:Ko-low} employing the Euler method with the same time step size of $0.1$. The observed trajectories \(R_{\mathrm{obs}}(t)\) and \(\Phi_{\mathrm{obs}}(t)\) were produced by simulating the HF model (Eq.~\eqref{eq:Ko-high}) at parameter values \(K=2\), \(\omega_0 = \pi/3\), and \(\gamma=0.1\). The discrepancy between the simulated and observed data is quantified by the metric 
\[
\Delta\big((R, \Phi), (R_{\mathrm{obs}}, \Phi_{\mathrm{obs}})\big) = \left\| S\big((R, \Phi)\big) - S\big((R_{\mathrm{obs}}, \Phi_{\mathrm{obs}})\big) \right\|_2^2,
\]
where \(S(\cdot)\) denotes the vector of summary statistics.

The observed data and the simulations generated from the LF and HF models at the true parameter values are shown in  Figure S.6 of Appendix Section S.2.2. Note that the LF simulations are deterministic, while the HF simulations incorporate stochasticity. A total of $5,000$ HF simulations required $55.75$ seconds, whereas $5,000$ LF simulations took only 2.017 seconds.

\begin{figure}[H]
    \centering
    \includegraphics[width=\linewidth]{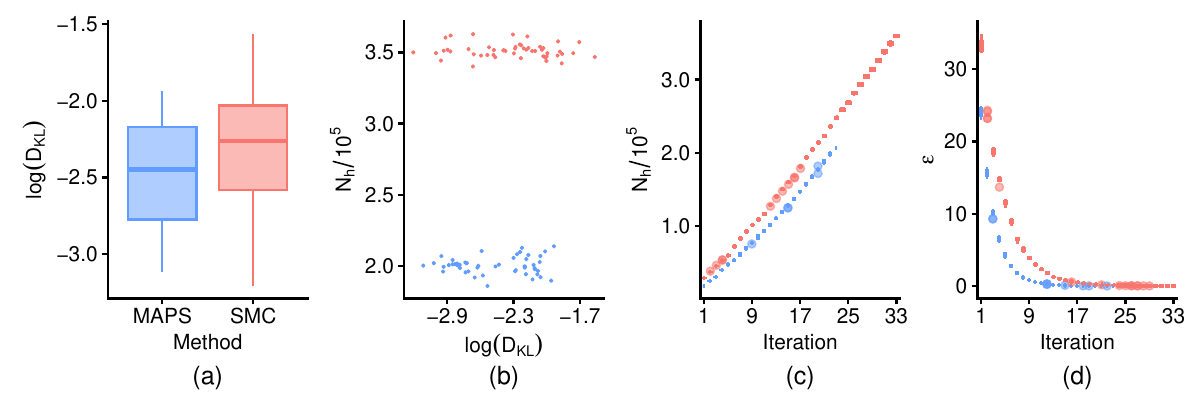}
    \caption{Comparison of results from 50 repetitions of ABC-ASMC and MAPS methods.}
    \label{fig:Ku_summary}
\end{figure}

Figure \ref{fig:Ku_summary} compares the performance of the MAPS and ABC-ASMC algorithms, each employing 2048 particles and repeated 50 times to ensure statistical robustness. For the MAPS algorithm, the number of simulations per parameter is set to \(n_H = 5\) for the HF model and \(n_L = 1\) for the LF model, with threshold reduction parameters \(\alpha_H = \alpha_L = 0.9\) and a baseline threshold \(a_L = 0.001\). The standard ABC-ASMC algorithm uses \(n = n_H = 10\) simulations per parameter with a threshold reduction parameter \(\alpha = 0.9\). The true ABC posterior is approximated via ABC-MCMC using 10 chains, each with 500,000 iterations, where the first 200,000 iterations are discarded as burn-in. Appendix Figure S.7 and S.8 show the evolution of bivariate marginal posterior densities across iterations, indicating faster convergence for MAPS.

\begin{figure}[H]
    \centering
    \begin{subfigure}[b]{0.24\linewidth}
        \centering
        \includegraphics[width=\linewidth]{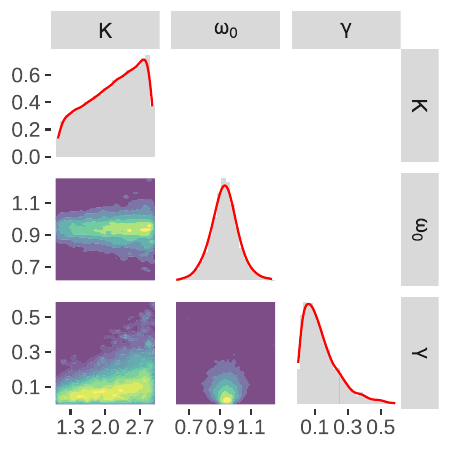}
        \caption{True posterior}
        \label{fig:KU_true}
    \end{subfigure}
    \hfill
    \begin{subfigure}[b]{0.24\linewidth}
        \centering
        \includegraphics[width=\linewidth]{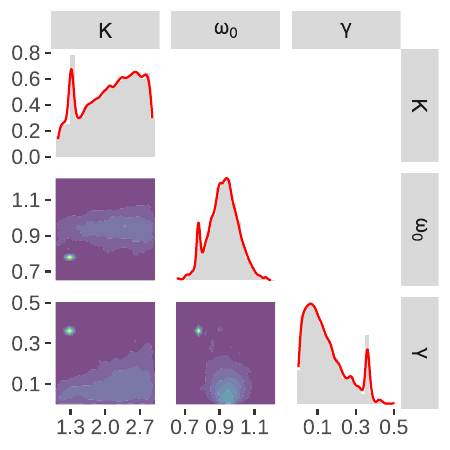}
        \caption{ABC-MCMC}
        \label{fig:KU_MCMC}
    \end{subfigure}
    \hfill
    \begin{subfigure}[b]{0.24\linewidth}
        \centering
        \includegraphics[width=\linewidth]{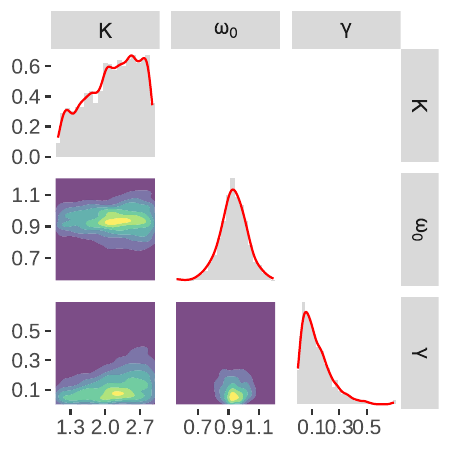}
        \caption{ABC-ASMC}
        \label{fig:KU_SMC}
    \end{subfigure}
    \hfill
    \begin{subfigure}[b]{0.24\linewidth}
        \centering
        \includegraphics[width=\linewidth]{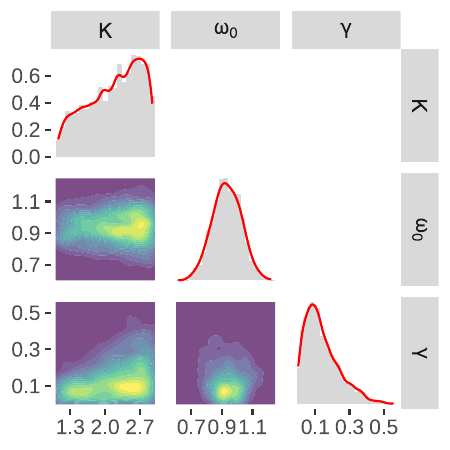}
        \caption{MAPS}
        \label{fig:KU_MAPS}
    \end{subfigure}
    \caption{Posterior estimates obtained by different methods.}
    \label{fig:KU_plot2D}
\end{figure}

In Figure \ref{fig:KU_plot2D}, we present the posterior distributions obtained by the ABC-ASMC, the MAPS methods, the ABC-MCMC algorithm, and the true posterior distribution. The ABC-MCMC results are based on $200,000$ iterations, with $10$ simulations per parameter, resulting in a total number of HF simulations comparable to that of the MAPS algorithm. The first $100,000$ iterations were discarded as burn-in. Compared to the SMC-based methods, the ABC-MCMC algorithm tends to be trapped in local modes, which negatively impacts its convergence.

\begin{figure}[H]
    \centering
    \includegraphics[width=1\linewidth]{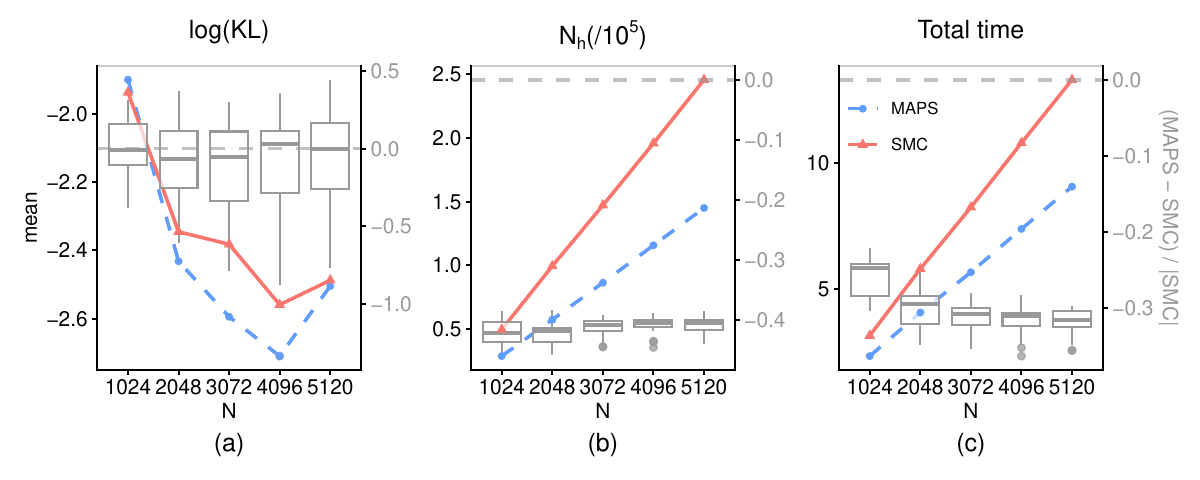}
    \caption{
Comparison of MAPS and SMC as function of \(N\) with \(\alpha=0.8\) (20 runs). Lines show the mean values, and boxplots indicate the relative performance of MAPS versus SMC.}
    \label{fig:KU_N_alpha0.8}
\end{figure}
\begin{figure}[H]
    \centering
    \includegraphics[width=1\linewidth]{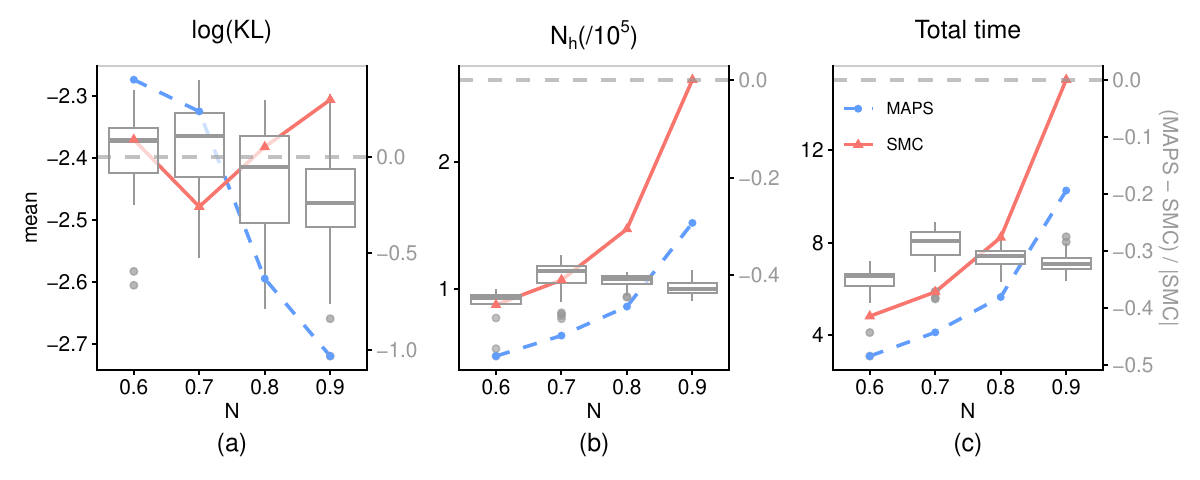}
    \caption{Comparison of MAPS and SMC as a function of $\alpha$ with $N=3072$ (20 runs). Lines show the mean values, and boxplots indicate the relative performance of MAPS versus SMC.}
    \label{fig:KU_alpha_N3072}
\end{figure}

Furthermore, we investigated the effects of the number of particles per iteration (\(N\)) and the threshold reduction parameter (\(\alpha\)) on the SMC inference. When fixing the threshold reduction parameter at \(\alpha=0.8\), Figure \ref{fig:KU_N_alpha0.8} shows that as \(N\) increases, the inference accuracy of both the ABC-ASMC and MAPS algorithms improves, while the cost of HF simulations and the total computational time grow linearly. When fixing \(N=3072\), Figure \ref{fig:KU_alpha_N3072} demonstrates that as \(\alpha\) increases, both the HF simulation cost and computational time of ABC-ASMC and MAPS increase;  the inference accuracy of MAPS improves, whereas ABC-ASMC shows no clear trend. 
The boxplots of Figure \ref{fig:KU_N_alpha0.8} and \ref{fig:KU_alpha_N3072} 
show the relative performance of MAPS versus SMC.
The results indicate that MAPS achieves inference accuracy comparable to or better than ABC-ASMC, while reducing HF simulation cost by approximately 40\% and total computation time by about 30\%.
\section{Discussion}
\label{sec: con}
We consider approximate Bayesian inference for simulation expensive models. 
Multifidelity simulation models are integrated in ABC algorithms and we propose a pre-filtering hierarchical importance sampling algorithm that leverages low-cost LF simulations to reduce the number of expensive HF simulations. We also present a strategy to assess the suitability of a given multifidelity model for our algorithm. Building upon the multifidelity model, we develop the MAPS algorithm that adaptively selects the pre-filtering criteria. We establish the posterior concentration property of the proposed algorithm under mild regularity conditions and prove that the hyperparameter $a_L$
governs the upper bound of  inference error. Numerical experiments corroborate these theoretical findings and show that our approach maintains accuracy comparible with standard ASMC while achieving approximately 40\% reduction in HF model evaluations. Moreover, the proposed method is applicable to both models with stochastic LF simulations (Sections \ref{sec:toy} and \ref{sec:OU}) and models with deterministic LF simulations (Section \ref{sec:KU}).

The MAPS algorithm introduces three parameters related to the pre-filtering stage: the number of LF simulations per parameter \( n_L \), the decay factor of the auxiliary threshold \( \tilde{\varepsilon} \), denoted by \( \alpha_L \), and the factor controlling the auxiliary threshold boundary \( a_L \). The selection of hyperparameters may impact the algorithmic performance in practice. We recommend choosing the decay factor \( \alpha_L \) to be consistent with the scale of the main threshold \( \varepsilon \), and setting \( a_L \) to 0.001. Regarding \( n_L \), the selection can be guided by the relative simulation costs: for each parameter, the cost of generating \( n_L \) LF simulations should not exceed 0.4 times the cost of conducting \( n_H \) HF simulations. 

Our proposed method differs from the multifidelity ABC algorithms of \cite{prescott2020multifidelity, prescott2021multifidelity} in several key aspects. Firstly, \cite{prescott2020multifidelity} constructs an unbiased estimator of the likelihood by combining low- and HF data, and based on this, designs early acceptance and early rejection algorithms that reduce the number of HF simulations.
Our method is based on using LF models to filter out parameters whose weights are highly likely to be zero. Building upon this, \cite{prescott2021multifidelity} proposes an MF-ABC SMC algorithm that relies on a fixed sequence of thresholds and employs kernel density estimation to approximate the current sample distribution as the importance proposal for the next intermediate target. In contrast, our method adaptively decreases the threshold by controlling the number of active particles. Particle weights are then updated accordingly, and a combination of resampling and MCMC propagation is used to generate weighted samples from the subsequent intermediate target distribution. The MF-ABC method of \cite{prescott2021multifidelity} is theoretically rigorous and demonstrated effectiveness, in practice it may produce samples with negative weights. 

There are several improvements and extensions as future work. The proposed method in this paper does not involve the development of LF models, which is assumed to be known a priori.  
 However, in some real applications, deriving a LF model may not be straightforward. 
 In future, it is worth exploring the construction of LF models using machine learning approaches and combining them with approximate Bayesian computation for inference. Multi-fidelity inference methods in the literature often rely on a single LF model to assist inference \citep{prescott2020multifidelity, cao2025using}.
 However, there is a trade-off between accuracy and computational speed. Simpler models tend to be faster but has lower accuracy. Another future direction is to incorporate models of multiple fidelity levels into the approximate Bayesian inference framework.

\section*{Acknowledge}
This work was supported by the National Natural Science Foundation of China (12131001 and 12101333), the startup fund and the Institute Development Fund and the HPC Platform of ShanghaiTech University, the Fundamental Research Funds for the Central Universities, LPMC, and KLMDASR. The authorship is listed in alphabetic order. 

\section*{Disclosure Statement}
The authors report there are no competing interests to declare.

\bigskip
\begin{center}
{\large\bf SUPPLEMENTAL MATERIALS}
\end{center}

\begin{description}

\item[Appendix S.1:] The proofs of all theoretical results presented in the main text.

\item[Appendix S.2:] Some numerical results not shown in the main text.

\item[Appendix S.3:] 
Some description of algorithms not shown in the main text.

\item[Appendix S.4:] 
A summary of notations used in the paper.

\end{description}

\bibliographystyle{chicago}
\bibliography{Bibliography-MM-MC}
\def\spacingset#1{\renewcommand{\baselinestretch}%
{#1}\small\normalsize} \spacingset{1}


\end{document}